# Switchable tribology of ferroelectrics

Seongwoo Cho[1], Iaroslav Gaponenko[2,3], Kumara Cordero-Edwards[2], Jordi Barceló-Mercader[4], Irene Arias[4,5], Céline Lichtensteiger[2], Jiwon Yeom[1], Loïc Musy[2], Hyunji Kim[1], Gustau Catalan[6,7], Patrycja Paruch[2*] and Seungbum Hong[1,8*]

[1]Department of Materials Science and Engineering, Korea Advanced Institute of Science and Technology (KAIST); Daejeon 34141, Republic of Korea

[2]Department of Quantum Matter Physics, University of Geneva; 1211 Geneva, Switzerland

[3]G.W. Woodruff School of Mechanical Engineering, Georgia Institute of Technology; Atlanta, Georgia 30332, United States of America

[4]Laboratori de Càlcul Numèric, Universitat Politècnica de Catalunya; Barcelona 08034, Spain

[5]Centre Internacional de Mètodes Numèrics en Enginyeria (CIMNE); Barcelona 08034, Spain

[6]Institut Català de Nanociència i Nanotecnologia (ICN2), Campus Universitat Autonoma de Barcelona; Bellaterra 08193, Spain

[7]Institució Catalana de Recerca i Estudis Avançats (ICREA); Barcelona 08010, Spain

[8]KAIST Institute for NanoCentury (KINC), Korea Advanced Institute of Science and Technology (KAIST); Daejeon 34141, Republic of Korea

*Corresponding authors: patrycja.paruch@unige.ch, seungbum@kaist.ac.kr

**Abstract:** Artificially induced asymmetric tribological properties of ferroelectrics offer an alternative route to visualize and control ferroelectric domains. Here, we observe the switchable friction and wear behavior of ferroelectrics using a nanoscale scanning probe where down domains having lower friction coefficient than up domains can be used as smart masks as they show slower wear rate than up domains. This asymmetry is enabled by flexoelectrically coupled polarization in the up and down domains under a sufficiently high contact force. Moreover, we determine that this polarization-sensitive tribological asymmetry is universal across ferroelectrics with different chemical composition and crystalline symmetry. Finally, using this switchable tribology and multi-pass patterning with a domain-based dynamic smart mask, we demonstrate three-dimensional nanostructuring exploiting the asymmetric wear rates of up and down domains, which can, furthermore, be scaled up to technologically relevant (mm-cm) size. These findings establish that ferroelectrics are electrically tunable tribological materials at the nanoscale for versatile applications.

Ferroelectrics, in which the switchable electric polarization is coupled with mechanical deformation, have been extensively studied in view of their technological applications as sensors, actuators, energy harvesters, and memory devices (*1-4*). Fundamentally, these materials exhibit electromechanically coupled properties including piezoelectricity (*1*), flexoelectricity (*5*), and electrostriction (*6*), whose interplay can enrich critical opportunities in the field of condensed matter physics and functional materials engineering, but is only now beginning to be understood and controlled (*7-9*). In addition, oppositely polarized ferroelectric surfaces present different mechanical responses under inhomogeneous deformation as a consequence of the interaction between flexoelectricity and piezoelectricity (*10*), which could be harnessed for the mechanical reading of ferroelectric polarization (*11*). As we shall show in this article, increasing this flexoelectric contribution under highly inhomogeneous stress also has emergent consequences for coupled tribological properties—in particular, friction and wear coefficients—which in turn can be exploited for direct visualization of ferroelectric domains and extremely fine physical lithography without the need for masks or chemical reagents.

Precise control of ferroelectric surfaces as well as domain structures is essential for numerous applications such as data storage (*3, 4*) and electro-optic devices (*12*). However, there are no previous studies on simultaneous or continuous control of both ferroelectric domains and the surface morphology nanostructure. During the nanostructuring of materials, including ferroelectrics, patterns with desired size, shape, and periodicity are transferred to the target substrate, generally via an intermediate bridging process using masking, resist, imprint or local thermochemical interactions (*13-16*). In contrast, intrinsic properties of the substrate such as ferroelectric polarization are rarely employed as a marker for patterning, although we note the demonstration of selective deposition of functionalized nanoparticles (*17*) and chemical reaction rate difference (*18*) depending on the surface chemistry of ferroelectric domains. Nevertheless, there are no reports in which the ferroelectric polarization is used as a smart mask for nanoscale patterning based on differential and locally switchable *mechanical*, *as opposed to chemical*, etch rates.

In this study, we establish that such switchable, polarization-dependent mechanical etching is in fact possible. We demonstrate that the local friction and wear behavior of ferroelectrics is asymmetric and independent of surface chemistry under large strain gradient, and that this inherent tribological asymmetry enables facile and reversible control of friction and wear properties, which can be exploited for nano-lithographic patterning by simply "rubbing" the surface of a voltage-written ferroelectric. To prove our idea, we start from uniaxial ferroelectric $LiNbO_3$ single crystals as a simple and accessible model system, already widely used in electro-optic applications (*19*). We discover polarization-dependent asymmetric friction and surface wear in these materials by applying a sufficiently high mechanical force using a diamond atomic force microscopy (AFM) probe at the nanoscale or silica particles at the bulk scale. Furthermore, we confirm that this asymmetry does not originate from either electrostatic effects, or inhomogeneous defect distribution, but is linked to the competing vs. synergistic interplay of flexoelectric and ferroelectric polarization in oppositely oriented domains, which moreover leads to an anomalous, positive correlation between the hardness of the materials and its wear rate. Switching the ferroelectric domains by local electric field application should thus allow simultaneous and reversible control of the tribological responses (friction and wear) of the material, which in turn can be used to dynamically manipulate surface morphology nanostructures. We extend our findings to $LiNbO_3$ and $PbTiO_3$ thin films to establish the universality of the observed tribological asymmetry, allowing for more precise polarization-derived friction microscopy and lithography, including single-lattice wear of ferroelectrics with atomic terrace edge features. Finally, we demonstrate this approach as a top-down, chemical-free and resist/mask-less lithography technique, which can be potentially applied to the fabrication of three-dimensional (3D) and monolithic nanostructures when multi-pass switching and milling of the ferroelectric surface is implemented.

The experiments were performed on commercially available periodically poled $LiNbO_3$ (PPLN) single crystals, composed of alternating out-of-plane polarization domains, further described in fig. S1. We observed polarization-dependent tribology of the sample surface using single-crystalline conductive diamond probes (NM-TC, *Adama Innovations*), selected for their extreme hardness and stiffness, with relatively high contact loading force (5 µN) and scan rate (4.88 Hz, equivalently 146.48 µm/s). The down domains oriented into the sample plane are less heavily etched than the up domains oriented out of the sample plane, resulting in strongly asymmetric milling after multiple scans, as shown schematically in Fig. 1A.

Accompanying this wear asymmetry, even though the pristine surface presents a flat morphology (Fig. 1B), we observed a polarization-dependent friction contrast (Fig. 1C) during the milling scans, with higher friction in the up-oriented domains. After the 50$^{th}$ milling scan by the diamond probe, the surface clearly shows the height difference between up and down domains (Fig. 1D); however, as demonstrated by subsequent piezoresponse force microscopy (PFM) imaging (Fig. 1E) using the same probe, this asymmetric milling process does not affect the domain polarity, which remains stable throughout the millings. The 3D surface plot in Fig. 1F also shows a clearly visible height difference after 50 milling scans, as does the line profile across the domains during the last milling scan in Fig. 1G, demonstrating strong height and friction difference between up and down domains. The height and friction differences signals (Fig. 1H) oscillate because of the contact geometry difference between frame-up (scan from below to above) and frame-down (scan from above to below) during the continuous, repeated milling scans, but we note that height (down-up) and friction (up-down) differences are always positive. The height and friction differences are clearly correlated with increasing scan number and show three distinct regimes. In region 1, the height difference slightly increases with relatively stable friction difference. In region 2, it drastically increases with higher friction difference, and the slope of height difference (etch rate) approaches a maximum value with maximum friction difference. In region 3, the height difference finally starts to saturate with constant lower friction difference. The etch depth compared with the pristine background region is approximately 4.56 nm for up domain and 3.64 nm for down domain regions (fig. S3). In addition, cross-validation of wear asymmetry using scanning electron microscopy (SEM) further indicates the contrast arises from the height difference of etched domains, not from the polarization difference (fig. S4).

Crucially, we find that the polarization-dependent asymmetric mechanical wear varies significantly as a function of the loading force (figs. S6–8). As detailed in fig S6, below a threshold loading force (less than 5 μN), no significant etching is observed, whereas when the force is too high (20 μN), surface deterioration and material fracture rather than steady asymmetric etching can dominate the resulting topography (*20, 21*). At the optimum loading force range (5–10 μN), we observed asymmetric etching of domains with some deterioration of the friction contrast and wear rate with continuous milling because of wear debris attachment of the tip, as further discussed in figs. S10 and S11. However, increasing or decreasing the scan rate during milling has insignificant effects on the tribological asymmetry (fig. S12).

Beyond micro/nano-scale wear of ferroelectric single crystals using an AFM probe, we also demonstrate scalability of this process to technologically relevant large-scale patterning by simply polishing the whole crystal using silica nanoparticles that effectively act as millions of mobile AFM tips (Fig. 1I). Fig. 1J shows the digital photograph of the crystal after such polishing and the optical microscope (Fig. 1K) and SEM (Fig. 1L) images exhibit the periodic boundaries of asymmetrically etched surfaces at large scale. As long as the nanoparticles were comparable in size to a typical AFM probe, giving rise to similar strain gradients under applied force, we could clearly see that up domains were preferably etched, resulting in nanoscale trench structures over 9-millimeter square area of the sample as seen in the height and PFM phase (Fig. 1, M and N), which is consistent with wear results observed using diamond probes. We believe that combining a bulk poling process with nanoscale silica bead polishing will enable us to realize bulk scale chemical-free/maskless lithography.

To understand the fundamental microscopic mechanisms behind our observations, we need to consider the possible interplay between friction—the resistance to relative motion between two surfaces—and mechanical wear—the removal of surface atoms after rubbing two surfaces—which can be strongly correlated if the loading force is sufficiently high, so that the probe indents the crystal during scanning. Friction and wear are complex tribological phenomena in which contributions of several possible micro/nanoscopic mechanisms could lead to the observed asymmetry of these responses in ferroelectrics (*22*). Previous studies have already reported on the asymmetric lateral force microscopy signals of ferroelectric single crystals (*23, 24*), but at a few tens of nN loading force such asymmetry might come from the effects of screening charges or adsorbates on the surface. However, if the friction properties are indeed governed by screening charges (electrons or holes) or adsorbates (especially chemisorbed species), the asymmetry should disappear with continuous milling scans, as any surface species or asymmetric skin layers are gradually removed. This is not the case in our results. Rather we find that the asymmetry persists throughout the full cycle of continuous etching.

Beyond the surface electrochemistry mechanism described above, another possibility is that the asymmetry emerges as a result of the different mechanical properties of up and down domains induced by the flexoelectric field, generated by the non-uniform strain applied via the AFM tip. Although the *ferroelectric* remanent polarization is itself symmetric, the additional *flexoelectrically* driven contribution leads to a different effective polarization in the up and down domains. This strain-gradient induced polarization has been shown to produce asymmetric mechanical responses (*10, 25, 26*). At the same loading force in up and down domains, we should therefore expect a difference in contact area during the etching scan, as schematically illustrated in Fig. 2A. Because friction strongly depends on the real contact area (*27*), as does the mechanical wear rate, we consequently expect higher friction and wear in up domains than down domains (Fig. 2, B and C).

Alternatively, in the second possible mechanism shown schematically in Fig. 2D, tribological asymmetry can arise electrostatically if the loading force is sufficiently high (around a few micronewtons) to enable screening charges to be scraped off the ferroelectric surface (*20, 28*). The unscreened ferroelectric surface of up vs. down domains has opposite electrostatic potential and field, which would lead to an asymmetric electrostatic force between the tip and sample (*29*). Further, charge transfer during sliding of the tip across the opposite polarization surface could possibly induce asymmetric friction (*30*).

We first examined the role of flexoelectricity in the friction responses using a self-consistent approach that couples flexoelectricity and piezoelectricity (*31*) based on a linear continuum theory of piezoelectricity (*32*). To confirm that flexoelectricity can explain the observed tribological asymmetry between ferroelectric up and down domains, we performed finite element simulations to quantify the indentation depth and contact area depending on polarization direction in the presence or absence of flexoelectric coupling, using the Signorini-Hertz-Moreau model for contact (*33*). The AFM tip is idealized as a conical rigid indenter in contact with an ideally flat $LiNbO_3$ surface. Further details of the theoretical and computational model are described in Supplementary Materials.

Indentation with the conical diamond probe generates a spreading electric potential distribution in the $LiNbO_3$ crystal (fig. S16). Since flexoelectric polarization interacts differently with up and down domains, this electric potential distribution varies depending on the direction of out-of-plane polarization. Piezoelectric and flexoelectric polarization fields upon indentation (fig. S17) thus lead to different competing or synergistic combinations in up and down domains. This combined interaction results in systematically higher indentation depths and contact areas for up domains (Fig. 2, E and F) and the asymmetry disappears in the absence of flexoelectricity (fig. S18). Furthermore, the differences in the indentation depths and contact areas increase as flexoelectric coupling becomes stronger (fig. S19). Experimentally, we do not have access to contact area measurements. However, if we assume a linear relation between contact area and friction, as predicted by single asperity friction models (*34*), the ratio of contact areas between up and down domains should be equal to the ratio of their friction, which is an experimental observable. From the simulation, this ratio is found to be independent of the indentation force (Fig. 2G). Experimental values of the friction ratio in Fig. 2G, acquired with changing loading force (Data from fig. S6), are consistently in agreement with simulated values for a flexocoupling coefficient of 10 V (*5, 35*). Experimentally, at lower loading forces (e.g., 2.5 µN), contact conditions deviate from ideal indentation during scanning, and thus no agreement with the simulated results should be expected.

To further discriminate between the flexoelectric vs. electrostatic mechanisms, we conducted nanotribology experiments using a non-conductive single crystalline diamond probe (D300, *SCD Probes*) to test the strength of possible electrostatic contributions. The probe was demonstrated to be non-conductive from the *I–V* curve measurement shown in fig. S20, but was still equivalently hard and stiff for mechanical etching of ferroelectric PPLN. Despite the non-conductive nature of the probe, we observed the same asymmetric behavior, with up domains having higher friction than down domains, as seen in Fig. 2H even with one milling scan at a loading force of 6 µN and scan rate of 4.88 Hz. Height and PFM phase measured using a conductive probe (Fig. 2, I and J) still remain stable after mechanical scraping, confirming that the asymmetry originated from pure mechanical milling of the probe on the ferroelectric surface. This finding can explain why electrically disconnected and insulating silica beads can mechanically etch the PPLN surface over a large area exactly the same way as the conducting diamond probe does over a small area.

Moreover, to further investigate the role of surface electrostatics from the other extreme, we tested electrostatic effects on the friction and wear of PPLN when applying an electric field. If the observed asymmetric friction and wear arise from the electrostatic asymmetry of the surface, we would expect high voltage application during milling to significantly change the results. However, when we applied such high voltage, varying from

-150 V to 150 V to the tip during etching scans, we observed no notable change in either the magnitude of the friction signal, or the contrast between the up and down domains during milling (Fig. 2, K and L). Furthermore, the resulting topography following 10 scans performed with the same varying applied bias shows no significant height change, as can be seen in Fig. 2M. We do, however, observe ferroelectric switching from up to down domain orientation occurring just as the voltage was varied between -150 V and 150 V, as shown in Fig. 2N. We note that these switched regions are slightly higher in topography than the unswitched up domain (Fig. 2M), further indicating the switchable nature of the wear asymmetry. We therefore exclude the electrostatic effects as a dominant contribution to the observed asymmetric nanotribological response.

Taken together, our simulation and experimental observations including other possibilities described in supplementary materials—also supported by previous studies (*10, 36*)— suggest that the dominant mechanism at the origin of the asymmetrical tribology is flexoelectrically induced polarization and the consequent asymmetry of mechanical properties between oppositely polarized up and down domains. We note that this asymmetry implies an anomalous, positive correlation between the hardness of the domains and their etch rate, while normally high hardness materials are more resistant to wear as reported in the form of Ashby plot between hardness and wear coefficient (*37*), the flexoelectrically modulated mechanical properties mean that up domains have larger hardness (*10*) *yet* are more strongly etched due to lower stiffness. Furthermore, the tribological asymmetry is indeed switchable, where the electrically reversed down domains exhibit lower friction signals than the up domains (fig. S22).

Using our observation of asymmetric wear, we could fabricate diverse nanostructures on the ferroelectric single crystal, manipulating the domain configuration by applying electrical bias using a nanoscale probe, as shown in Fig. 3A. This also enables 3D tomographic studies of ferroelectric nanostructures with continuous etching scans to map the full evolution of the pillar structure during subsequent milling, as schematically depicted in Fig. 3A. To demonstrate, in the up domain of the initial PPLN configuration, we switched a 4 × 4 array of nanoscale circular down domains by applying 150 V to the stationary AFM tip on the ferroelectric surface at the designated array positions for 120 s. Since down domains are more resistant to mechanical etching than up domains, this domain pattern allowed us to create an array of nanopillars milled using a diamond probe. The corresponding height, PFM phase and PFM amplitude are detailed in fig. S23, while the height and PFM phase after ten milling scans with 5 μN at 1.95 Hz are shown in Fig. 3B and 3C. The pillar height saturates at approximately 14.13 nm around the 60$^{th}$ milling scan (Fig. 3D). We note that initially, and for a small number of milling scans, the wear process yielded well-defined low nanopillars with a cylinder-like structure and a flat top. However, as the milling continues, progressive wear also of the down-oriented domains led to a more cone-like morphology, with a rounded top as the full growth of the pillars can be seen in Fig. 3E.

The same lithographical concept can be applied to thin film ferroelectrics where more precise friction visualization and lithography techniques are possible, although the procedures may require greater care because decreased film thickness promotes stronger strain gradients and can lead to flexoelectric switching (*38*). We demonstrated that this polarization-determined lithography could be successfully reproduced at reduced crystal dimensions such as in a 100 nm $LiNbO_3$ thin film. Simply switching ferroelectric domains electrically and milling at an optimum loading force regime can create a desired nanostructure (fig. S24). Height, PFM phase and 3D surface images with color overlapped with their PFM phase in Fig. 3F–H exhibit that facile fabrication without any chemicals or photomask is realized after artificially decorating the thin film with the text "FERRO" with down domains, and "LITHO" with up domains before etching. Height and PFM phase line profiles (Fig. 3I) clearly show the down domains as higher than up domains in the etched region. To demonstrate the generality of this phenomenon, and its independence of chemical composition and crystal structure, in addition to uniaxial $LiNbO_3$ single crystal and thin film, we carried out polarization-derived lithography in tetragonal $PbTiO_3$ thin film ferroelectrics. Once again, we observe that locally switched down domains (Fig. 3J) have the lower friction signal (Fig. 3K). As shown in Fig. 3L, atomic $PbTiO_3$ terrace edges, which originate from the $SrTiO_3$ substrate (fig. S26), were revealed after multiple etching scans at gradually increased loading force (up to 1600 nN). Surprisingly, the height difference between etched up and down domains is around 4 Å, which implies the height difference is indeed one cell height calculated from the XRD data (4.144 Å).

Furthermore, reproducible wear rate control based on a choice of polarization orientation, which can also be configured during the milling process, allows for the fabrication of more complex 3D nanostructure. Fig. 4A shows a schematic of such a multistep lithographic process at different stages of fabrication. From the pristine domain state (up in this case), in which the desired domain nanostructure can be patterned by applying electric bias to the scanning probe, initial mechanical milling establishes the desired height difference between the down and up domains. Subsequent rewriting of the domain structure and additional mechanical milling can then be repeatedly alternated to create the target nanostructure with multiple height levels. A final switching step can then be carried out to obtain a uniform polarization orientation throughout the sample, depending on the purpose of the device; if the application requires stiffer surfaces, a homogeneous down orientation is recommended, whereas up orientation is suggested for more flexible surfaces.

A proof-of-concept example of this novel approach to developing complex 3D structures is shown in Fig. 4B–F. The complete evolution of the 3D structure during repeated writing and milling is shown in figs. S27 and S28. The down oriented rectangular domain and the text "FERRO LITHO" are first electrically patterned and milled before the up oriented schematic AFM probe is electrically patterned. The AFM tip feature is fabricated by additional etching scans after this switching step. The height and PFM phase (Fig. 4, B and C) show the 3D ferroelectric nanostructure after the $2^{nd}$ milling process. The height image in Fig. 4D, including both pristine and etched areas, indicates the different topography as does the line profiles in Fig. 4E in which the distinct 3D structure can be discerned. We note that the use of a nanoscale probe allows us to create complex structures of nanoscale ferroelectric domains, and therefore, the resulting mechanical lithography shows significant technological promise. Although many previous studies on scanning probe lithography successfully carried out sample structuring, with a height difference obtained via mechanical, thermal or chemical etching (*15, 16, 39*), the present work is the first to demonstrate scanning probe nanostructuring using the tribological asymmetry between domains with different polarization orientations.

To summarize, our work reveals the asymmetric friction and wear of ferroelectrics, which opens up an alternative way towards probing and manipulating ferroelectric domains based on the switchable tuning of their tribological properties. We determine that the higher friction and thus faster wear rate in up domains originate from the strain-gradient driven flexoelectric response which either competes with or enhances the ferroelectric polarization in oppositely oriented domains, resulting in higher friction in up domains than down domains. Furthermore, our findings enable us to propose a simple methodology for patterning a desired 3D structure with arbitrary complexity by alternating electrical switching and mechanical milling steps. Finally, we establish the universal nature of tribological asymmetry independently of chemical composition or crystal structure, and demonstrate single-lattice scale wear in epitaxially grown ferroelectric thin films. We envision that this top-down, chemical/resist-free and maskless lithography technique can be scalably applied in the fabrication of ferroelectric nano/microstructures.

**Acknowledgments:**

**Funding:** S.H. acknowledges the National Research Foundation of Korea (NRF) grant funded by the Korea government (MSIT) (Grant No. 2020R1A2C201207811) and the KAIST-funded Global Singularity Research Program for 2021 and 2022. P.P. acknowledges Division II of the Swiss National Science Foundation under project 200021_178782. S.C. acknowledges the National Research Foundation of Korea (NRF) grant funded by the Korea government (NRF-2018-Global Ph.D. Fellowship Program). I.A acknowledges the support of the European Research Council (No. StG-679451 to I.A.), the Spanish Ministry of Economy and Competitiveness (No. RTI2018-101662-B-I00), and the Generalitat de Catalunya (ICREA Academia award for excellence in research to I.A. and Grant No. 2017-SGR-1278 to I.A.). CIMNE is a Severo Ochoa Centre of Excellence (2019-2023) under the grant CEX2018-000797-S to I.A. funded by MCIN/AEI/10.13039/501100011033.

**Author contributions:** S.C., P.P. and S.H. conceived the idea of asymmetric milling. S.C., I.G. and K.C.-E. designed the experiments. S.C. performed milling and scanning probe measurements. J.B.-M. and I.A. conceived the computational study and analyzed the results. C.L. prepared the PTO thin film sample. S.C. analyzed the SPM data. J.Y. and L.M. supported the data analysis. S.C. and H.K. performed SEM measurements. All authors discussed the results and edited the manuscript.

**Competing interests:** Authors declare that they have no competing interests.

**Data and materials availability:** All data are available in the main text or the supplementary materials.

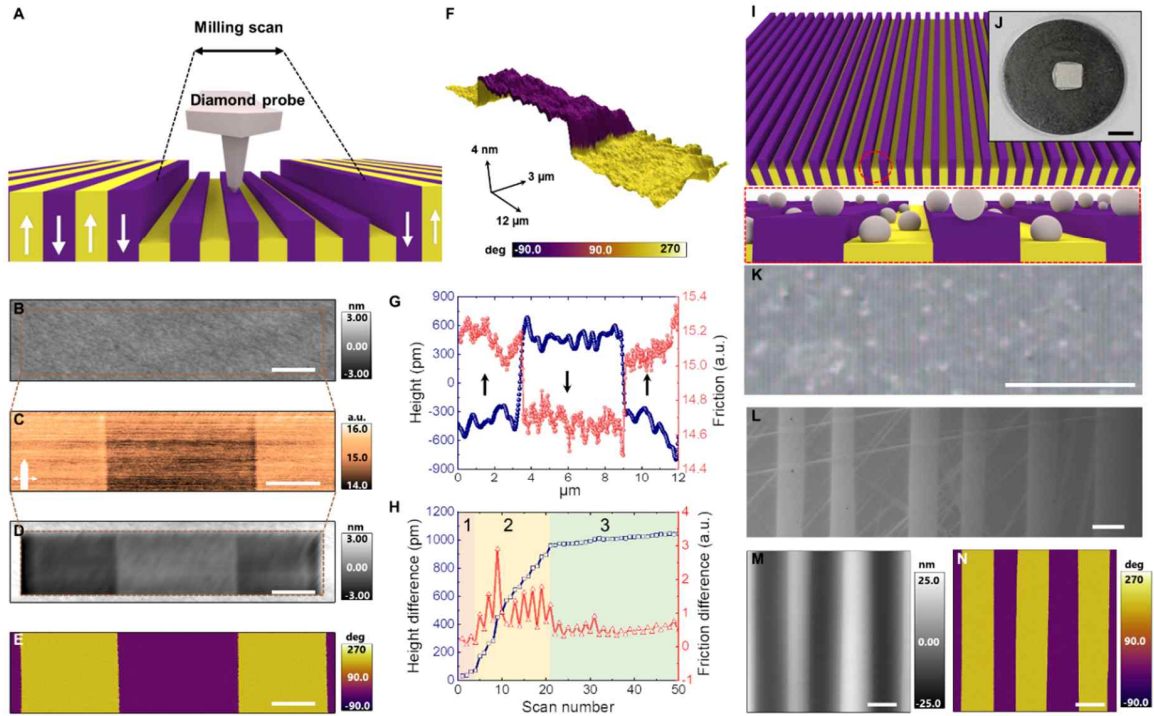

**Fig. 1. Observation of asymmetric friction and wear of ferroelectric LiNbO$_3$ single crystal.** **(A)** Schematic of asymmetric milling of PPLN using a diamond probe. Ferroelectric up and down domains exhibit different mechanical etch rates after milling without any external voltage applied to the probe. Pristine up and down domains begin with identical initial topography, but show different heights after repeated milling scans. **(B)** Height before milling, **(C)** friction image acquired during the 50$^{th}$ milling scan (scan angle of 90° with the fast scan axis perpendicular to domain walls). **(D)** Height and **(E)** PFM phase after fifty milling scans show stable domain orientation after the etching process. **(F)** 3D surface plot of etched surface after 50 milling scans with color scale indicating PFM phase. **(G)** Height and friction signal during the 50$^{th}$ milling scan. **(H)** Height and friction difference between up and down domains with increasing scan number. Scale bars in (B–E) are 2 μm. **(I)** Schematic of large area milling by simply polishing the crystal using silica particles. **(J)** Digital photograph of 3 × 3 mm$^2$ PPLN. **(K)** Optical microscope image and **(L)** scanning electron microscope image of etched PPLN. **(M)** Height and **(N)** PFM phase of large area milled PPLN. Scale bars are 3 mm for (J) 300 μm for (K) and 5 μm for (L–N).

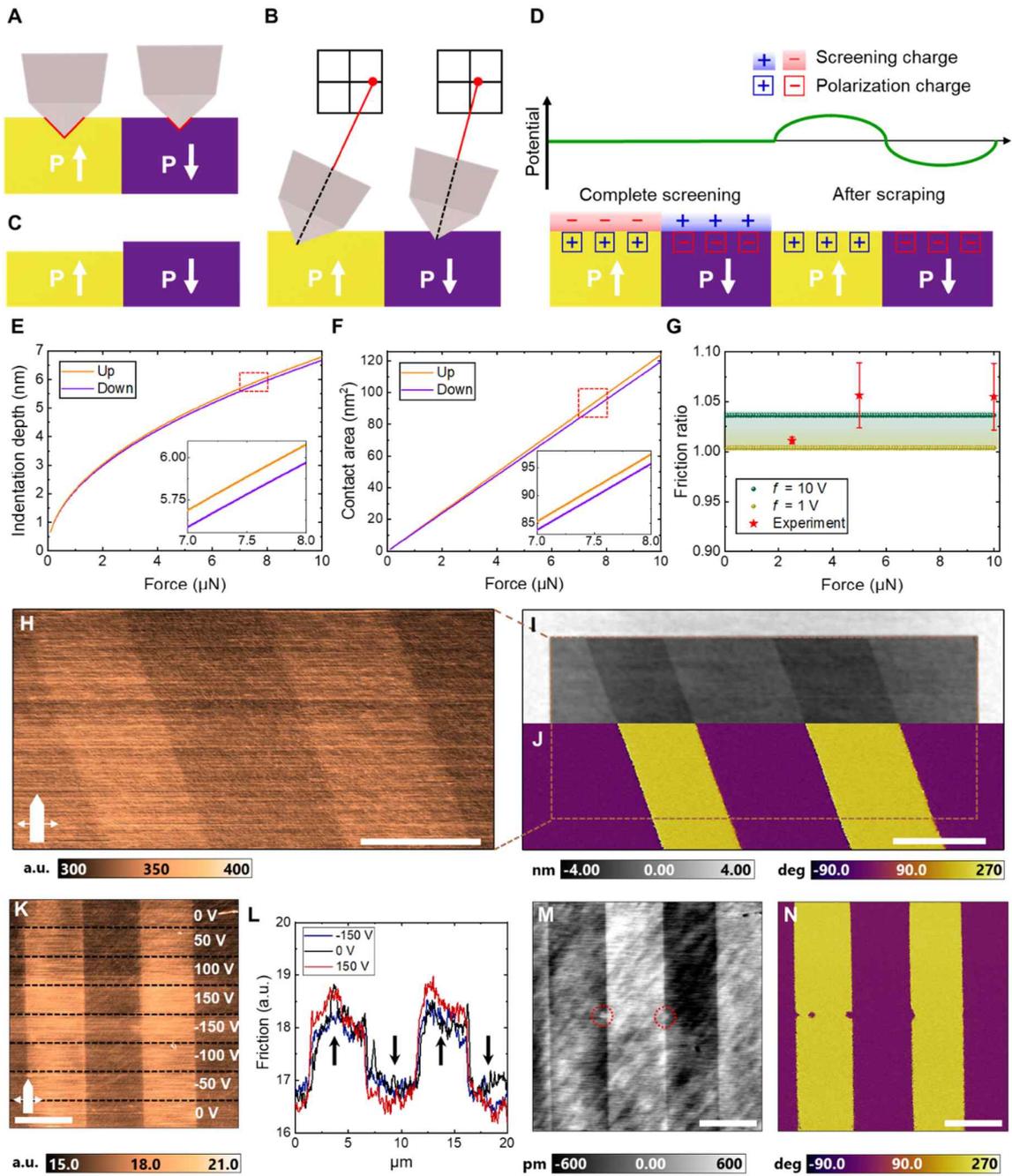

**Fig. 2. Origin of asymmetric friction and wear of ferroelectrics. (A–C)** Schematic of flexoelectrically induced mechanism, in which the asymmetric tribological properties depend on the direction of the out-of-plane polarization, leading to larger contact area during the scanning in up domains. **(A)** The probe sinks deeper in the ferroelectric up domain than down domain because of stiffness asymmetry. **(B)** During the dynamic scan, the absolute value of the lateral signal in position-sensitive photodiode is higher in the up domain than down domain because of larger contact area and indentation depth. **(C)** Schematic of unscreened surface charge driven mechanism. **(D)** The resulting height is higher in the down domain due to continuous milling with higher friction in the up domain. **(E–G)** Computational mechanics approach of asymmetric friction in $LiNbO_3$. **(E)** Indentation depth and **(F)** contact area of up and down domains based on cubic flexoelectricity with equal longitudinal and transversal coefficients corresponding to a flexocoupling coefficient of 10 V. **(G)** Friction ratio (up/down) obtained from the numerical simulation and experiments. Flexocoupling coefficients simulated in (G) are between 1 and 10 V. **(H–J)** Investigation of tribological asymmetry with a non-conductive diamond probe. **(H)** Friction during the milling scan with non-conductive probe, **(I)** resulting topography (etched inside) with pristine background region and **(J)** PFM phase using conductive diamond probe after one milling scan. **(K–N)** Results of high voltage application to tip during the milling. **(K)** Friction during the 1st milling scan and **(L)** line profiles in 150 V, 0 V and -150 V region. **(M)** Resulting topography and **(N)** PFM phase after 10 milling scans (scan angle of 90° with the fast scan axis perpendicular to the domain walls). Scale bars in (H–N) are 5 μm.

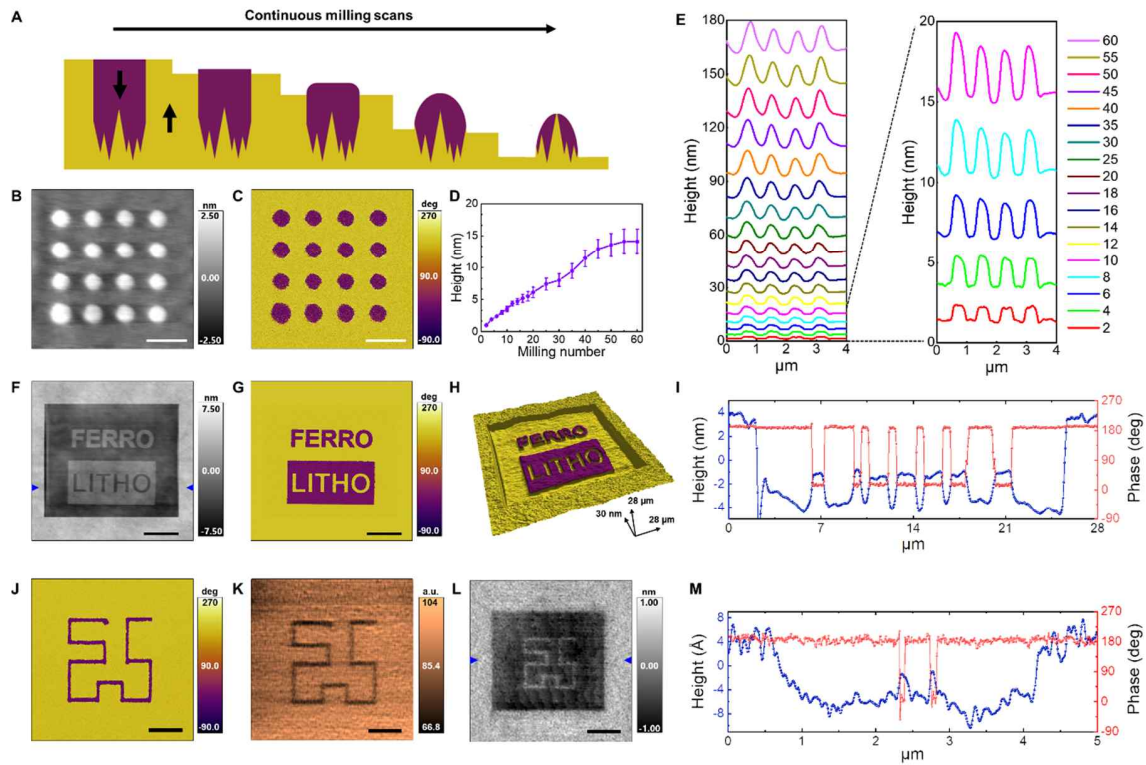

**Fig. 3. Universal tribological asymmetry.** The tribological asymmetry works universally regardless of size and type of ferroelectric. **(A–E)** Nanopillar fabrication on PPLN single crystal by continuous mechanical milling after electrical switching, both with the same conductive diamond probe. **(A)** Schematic of nanopillar fabrication on PPLN. **(B)** Height and **(C)** PFM phase after ten milling scans at a loading force of 5 μN and a scan rate of 1.95 Hz. Scale bars in (B) and (C) are 1 μm. **(D)** Average pillar height evolution with continuous milling scans. **(E)** Growth of nanopillars under continuous milling scans based on line profiles from four pillars. **(F–I)** Ferroelectric nanostructure fabrication on thin $LiNbO_3$ film. **(F)** Height, **(G)** PFM phase and **(H)** 3D surface images with color overlapped with their PFM phase. Scale bars in (F) and (G) are 6 μm. **(I)** Height and PFM phase line profile along the blue marker in (F). **(J–M)** Nanofabrication of $PbTiO_3$ thin film. **(J)** PFM phase after the artificial decoration of ferroelectric domains, **(K)** Friction image during milling scans at 800 nN and **(L)** height after multiple milling scans. Scale bars are 600 nm for (J) and (K), and 1 μm for (L). **(M)** Height and PFM phase line profile along the blue marker in (L).

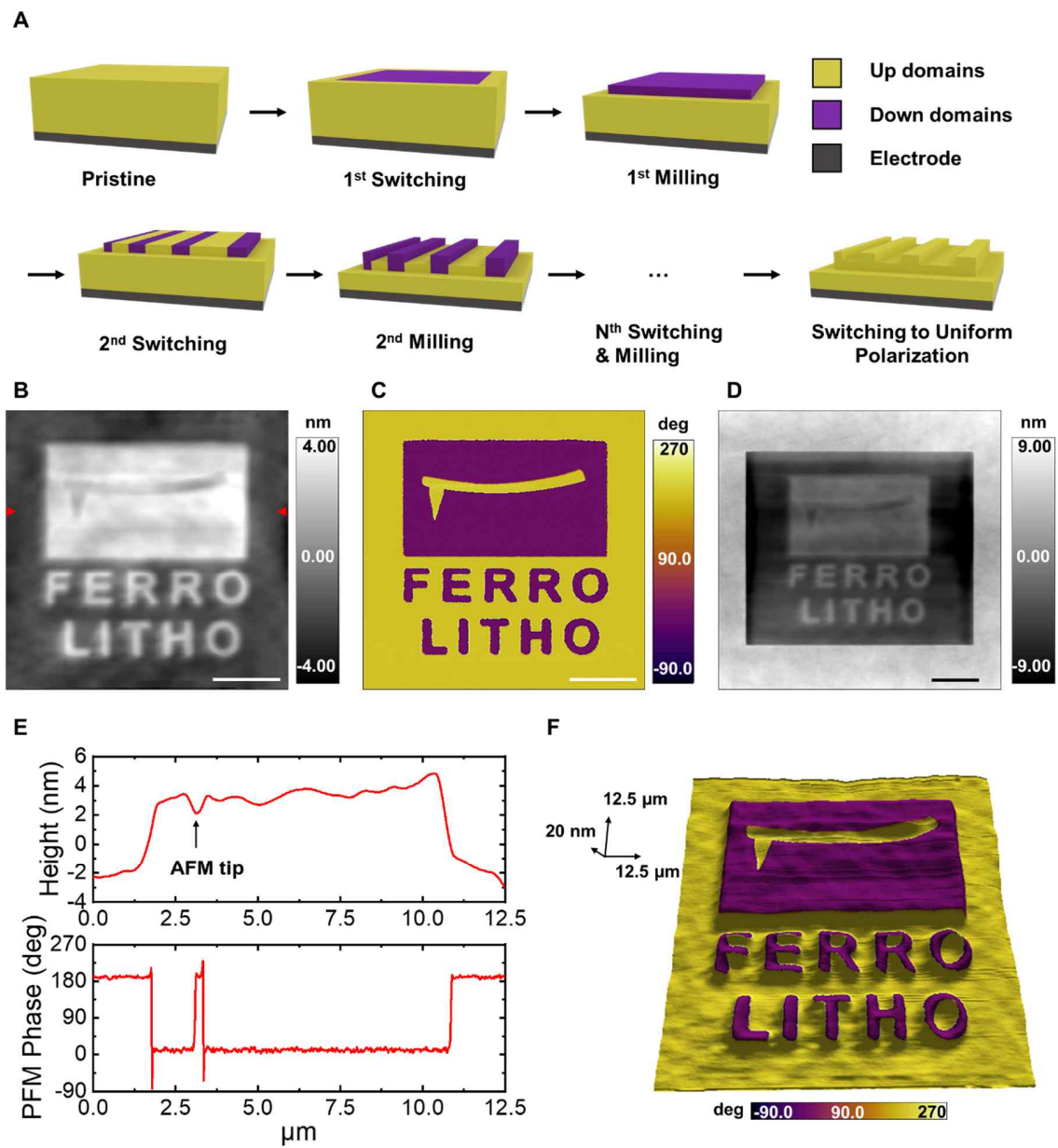

**Fig. 4. 3D nanostructure fabrication using asymmetric nanotribology on ferroelectric LiNbO$_3$ thin film. (A)** Schematic of top-down, chemical-free and maskless multi-patterning combining the switchable nature and asymmetric wear of ferroelectrics. **(B)** Height and **(C)** PFM phase after multiple switching and etching steps. **(D)** Height, including pristine background area, after multiple patterning. **(E)** Line profiles of height and PFM phase along the AFM probe feature in (B). **(F)** 3D representation of the multi-patterned structure with PFM phase color superimposed on the height image.

**Supplementary Materials**

**Materials and Methods**

Materials

Three different LiNbO$_3$ samples were used to probe the asymmetric friction and etching. A periodically poled lithium niobate (PPLN) single crystal (AR-PPLN test sample, 3 × 3 mm$^2$ with a thickness of 500 μm, *Asylum Research*) was chosen for the primary experiments demonstrating the polarization-dependent asymmetric tribology from Figs. 1, 2, 3A–E, S1–13, S22 and S23. A stoichiometric LiNbO$_3$ (optical grade single crystal, 10 × 10 mm$^2$, with a thickness of 500 μm, *MTI Corporation*) was used in control experiments to investigate the possible effects of variations in defect density (Fig. S21). For prototype nanostructure fabrication (Fig. 3F–I, Fig. 4 and Figs. S24–29), 100 nm thick z-cut undoped LiNbO$_3$ films on Cr/SiO$_2$/LiNbO$_3$ substrates with z+ oriented polarization (*NanoLN*) were prepared by cutting from single crystal congruent LiNbO$_3$ using the ion-slicing method (*40*).

The PbTiO$_3$ thin film in this study was grown epitaxially using off-axis radio frequency magneton sputtering on a TiO$_2$ terminated, (001)-oriented SrTiO$_3$ substrate. A SrRuO$_3$ layer was first deposited at 640°C in 100 mTorr of O$_2$/Ar mixture of ratio 3:60 using a power of 80 W, using a stoichiometric target. The PbTiO$_3$ film was then grown in-situ in 180 mTorr of a 20:29 O$_2$/Ar mixture using a power of 60 W at a temperature of 540 °C using a Pb$_{1.1}$TiO$_3$ target with a 10% excess of Pb to compensate for Pb volatility. The SrRuO$_3$ layer is 60 $\pm$ 5 unit cells thick, with a c-axis = 3.975 $\pm$ 0.002 Å. The PbTiO$_3$ layer is 210 $\pm$ 8 unit cells thick, with c-axis =4.144 $\pm$ 0.001 Å. The as-grown sample is atomically flat, with a roughness of the order of 2 Å (Fig. S25).

Polarization-derived friction microscopy and lithography

All scanning probe measurements were performed using a commercial atomic force microscope (Cypher ES and Cypher VRS, *Asylum Research*). Single-crystalline conductive diamond tips (NM-TC, *Adama Innovations,* Lot number: 009-013) were used to image the surface friction response and to etch the LiNbO$_3$ samples. Ferroelectric LiNbO$_3$ nanopillars were fabricated using a single crystalline diamond probe (NM-TC, *Adama Innovations,* Lot number: 009-013), where electrical switching and mechanical milling were conducted using the same diamond probe. Au-coated Si probes (4XC-GG, *MikroMasch*) were used for visualization and lithography of $PbTiO_3$ thin film to minimize the effect of flexoelectric switching. After the milling scans of $PbTiO_3$ thin films, pristine probe was used for higher resolution. During the high force contact imaging, DC or AC electric bias was not applied through the probe except the simultaneous piezoresponse force microscopy (PFM) imaging at high contact force in Fig. S9. The loading force during the scan was calculated by multiplying the inverse optical lever sensitivity obtained from the force-distance curve measurements, the spring constant of the probe, and set point difference in the position-sensitive photodiode in contact with the sample. The scan angle was fixed at 90° (perpendicular to the long axis of the cantilever) during friction imaging and milling.

PFM was performed at to image polarization domains at a loading force much less than the milling force to prevent sample damage during the imaging except the case to check the possible transient switching during high force application in Fig. S9. During some of measurements, temperature and humidity were maintained at set levels in the environmental cell with an in-house-designed low-noise humidity controller to ensure the stable condition of the ferroelectric surface (*41, 42*). During the switching process in the nanostructuring experiments in $LiNbO_3$ thin film, Pt coated Si probes (HQ:DPER-XSC11, *MikroMasch*) were used to switch the rectangular and "FERRO LITHO" domain from up to down in the first switching to prevent tip contamination before milling scans. After the first switching, milling scans were conducted under the optimum conditions using diamond probes. The same diamond probe was used in the second switching, creating an AFM probe feature in the rectangular down domain, following multiple milling scans and switching to obtain a uniform polarization state.

Large scale polishing of ferroelectric single crystals

PPLN single crystals were gently milled and polished using Multiprep$^{TM}$ Polishing System (*Allied*). Following mechanical grinding of the crystal using diamond lapping films (1, 3, 9, and 15 μm, *Allied*), final polishing was performed with colloidal silica suspension (0.04 μm, *Allied*) on a synthetic polishing cloth (Vel-Cloth, *Allied*). During polishing, the platen speed was 150 rpm at 2.94 N (300 gF) for 3 min. As control experiments, three different samples were prepared; We conducted mechanical grinding of the crystal using diamond lapping films, dipping in the colloidal silica solution and polishing using colloidal suspension, separately in three pristine PPLN (Fig. S13).

Flexoelectrically-coupled contact mechanics simulations

The complete model of flexoelectrically-coupled contact mechanics is described in the computational model and contact model sections in the supplementary materials. We used the model to compute the indentation depth and contact radius of a conical diamond tip indenting on a poled LiNbO$_3$ sample accounting for flexoelectricity. A rectangular computational domain is considered, with $L = 40$ nm and $H = 20$ nm. The material parameters for elasticity, dielectricity and piezoelectricity are taken from an open database of computed material properties (*43*). We considered a simplified form of the strain gradient elasticity tensor $h_{ijklmn} = (\lambda \delta_{ij} \delta_{lm} + 2\mu \delta_{il} \delta_{jm})\ell^2 \delta_{kn}$, where $\lambda$ and $\mu$ are Lamé elasticity parameters, $\ell$ is the length-scale taken as 10 nm, and $\delta_{kn}$ is the Kronecker delta. We assumed a cubic isotropic form of the flexoelectric tensor, with two independent components, longitudinal $\mu_L$ and transversal $\mu_T$, the shear component being $\mu_S = \frac{1}{2}(\mu_L - \mu_T)$ (*44*). In all simulations in this work, we took $\mu_L = \mu_T = \varepsilon_0 \varepsilon_r f$, where $f$ is the flexocoupling coefficient; thus, $\mu_S = 0$. The 3D solution of the axisymmetric problem is shown in Fig. S16. considering an upward polarized sample, for $f = 10\ V$. The boundary conditions of the corresponding axisymmetric problem are as follows: left side clamped horizontally and bottom side grounded and clamped vertically.

Estimation of friction force

To quantitatively compare the simulated asymmetry in the indentation depth to that observed experimentally in the friction, we assumed a linear relation between the friction force and contact area (*34, 45, 46*). Consequently, the ratio between contact areas in up and down domains, should be equal to that of their friction. For a conical indenter, we can write $F_f = \tau A_c = \tau \pi R_c \sqrt{R_c^2 + d^2}$ where $F_f$ is the friction force, $\tau$ is the shear strength, $A_c$ is the contact area, $R_c$ is the contact radius and $d$ is the indentation depth. The ratio $\frac{F_f^{up}}{F_f^{down}} = \frac{A_c^{up}}{A_c^{down}}$ is given in Table S2 for different values of flexocoupling coefficient.

Periodically poled lithium niobate (PPLN)

Asymmetric friction and wear were observed in the z-cut PPLN single crystal. LiNbO$_3$ is a simple and ideal model system for probing asymmetric tribology because ferroelastic switching (e.g., switching from the c-domain to the a-domain or vice versa) is limited. Therefore, only up and down domains exist, even with a high strain gradient. Without the possible effects from the a-domain in the crystal (*47*), we investigated PPLN single crystal with the schematic in Fig. S1A. Topography, PFM amplitude and phase were visualized by single frequency PFM imaging using a conductive diamond probe (NM-TC, *Adama Innovations,* Lot number: 009-013). Fig. S1B shows the flat surface of the pristine PPLN (roughness = 251.2 pm). PFM amplitude and phase (Figs. S1C and S1D) indicate uniaxial out-of-plane ferroelectric domains (up and down), where the domain width is approximately 5 μm with periodic stripes.

Effect of loading force on the asymmetric nano-tribology

With a pristine single-crystalline conductive diamond tip, we optimized the asymmetric etching in four different regions with the same polarity. We scanned each area 10 times with loading forces from 2.5 to 20 μN, as shown in Fig. S6A. Fig. S6A shows the height after ten scans performed with different loading forces, and Fig. S6B shows the PFM phase image, confirming the stability of domain configuration after milling scans. We found an optimal loading force regime for the asymmetric etching. If the loading force is low (i.e., below 2.5 μN), no notable etching of the crystal is observed (the etch depth calculated by subtracting the average height value of the milled surface from the pristine background surface is -16.0 pm, which is much lower than sample roughness.) If the loading force is in the optimum regime (from 5 to 10 μN), asymmetric etching occurs, and the up and down domains have different heights, with higher average friction in up domains (Friction$^{up}$/Friction$^{down}$ is ~ 1.057 at 5 μN and ~ 1.051 at 10 μN). In the 10 μN case, the left side experiences partial breaking, which starts from the overscanning region, indicating a higher loading force applied in this overscanned region. If the loading force is high (i.e., approximately 20 μN), the resulting topography is dominated by etching with material fracture rather than asymmetric etching. Therefore, the loading force should be higher than the force required to initiate the etching but lower than the material fracture regime for asymmetric etching.

Figs. S7 and S8 also depict the effects of increasing and decreasing loading force on friction asymmetry and wear behavior. With a pristine diamond probe, each region was milled four times with a loading force starting from 1 to 10 µN in 10 different regions with the same polarity (Fig. S7A). We observe no significant etching in the regions milled at 1 to 3 µN, but do observe asymmetric etching in regions milled from 4 to 10 µN. Fig. S7B shows the height image obtained in an analogous manner but this time decreasing the loading force from 10 to 1 µN using the same tip, and we observe asymmetric etching between 10 and 7 µN. Fig. S7C shows the PFM phase in the region in Fig. S7B before millings, and it indicates alternating up and down domains. We believe that the origin of this hysteretic behavior is the probe degradation, which is further described in Figs. S10 and S11.

The etched height of PPLN is also seen in the line profiles in Figs. S8A and S8B. In significantly etched regions (4–10 µN in Fig. S7A and 7–10 µN in Fig. S7B), the etch depth of the down domain (purple) is always lower than that of up domains (orange). Figs. S8C and S8D indicate friction signals versus scan numbers with height during specific scans. As can be seen in the height images in Fig. S8C, the asymmetric etching starts during the 2$^{nd}$ scan at 4 µN. We observe the friction signals jump up from 2$^{nd}$ scan at 4 µN because of higher resistance to the AFM tip motion during etching. In Fig. S8D, the asymmetric etching starts from the 3$^{rd}$ scan at 7 µN, and the friction signals also rapidly increase at the 3$^{rd}$ scan at 7 µN, which implies a strong correlation between friction and asymmetric etching behavior.

Further, no transient flexoelectric switching occurred during the high force application as can be seen in Fig. S9. We simultaneously visualized ferroelectric domains with increasing loading forces from 200 nN to 20 µN, but we observe no flexoelectric switching from up to down domain even with loading forces (e.g., 15 and 20 µN) in material fracture regime.

Degradation of friction and wear

The asymmetric friction and wear rate show degradation with continuous milling scans, as can be seen in Fig. S10. Fig. S10A shows the PFM phase image before milling scans. Fig. S10B shows the resulting height after 20 milling scans in each region from region 1 (R1) to region 4 (R4), with the same milling conditions (loading force of 5 µN, scan rate of 1.95 Hz, controlled environmental condition of 20°C and relative humidity of 20%). Even under the same conditions, the average height difference in a pristine area shows a 61.1% decrease from R1 (96.5 nm) to R4 (37.5 nm). Figs. S10C and S10D show the etch depth in each region and especially the height difference between up (orange) and down domains (purple) in Fig. S10D. Fig. S10E shows the degrading behavior of friction signals, where the first friction signal in each region showing relatively lower friction might originate from the skin layer or adsorbates on the surface during the first milling scan. The average asymmetric friction tends to deteriorate with scanning as demonstrated by the signal in each region, which reveals a 35.9% decrease from R1 to R4. The correlation between the average friction and wear in each region shows a linear degradation from R1 to R4, as shown in Fig. S10F, suggesting that the mechanism possibly originated from asymmetric mechanical stiffness differences in up and down domains, because the quality of mechanical contact determines the etching efficiency. Furthermore, this correlation implies that the optimal loading force for the surface milling changes with continuous etching. The scanning electron microscopy (SEM) images of the pristine tip and the same tip after etching (Fig. S11) exhibit clear differences, with wear debris attachment on the tip side after milling scan making non-ideal indentation between the tip and the sample.

Effect of scan rate on the asymmetric nano-tribology

The effect of the scan rate on nano-tribology was investigated in the PPLN sample using a pristine single crystalline conductive diamond tip (NM-TC, *Adama Innovations,* Lot number: 009-013). Four different regions with the same polarity in each image were milled 10 times. First, the scan rate is increased from 1.26 to 9.77 Hz in four regions with a loading force of 5 μN, as can be seen in Fig. S12A. The wear depth decreases with increasing scan rate. However, using the same probe immediately after the experiment in Fig. S12A, with decreasing scan rate from 9.77 to 1.26 Hz at the loading force of 5 μN, we observe drastically reduced wear depth rather than the effect of scan rate on the wear depth. We then increased the loading force from 5 to 8 μN with decreasing order of scan rate from 9.77 to 1.26 Hz, and again observed degrading behavior, concluding that the tribological asymmetry is dominated by a change in the loading force and continuous degradation during the etching, rather than a change in scan rate. PFM phase images after three sets of experiments show stable domain configuration after experiments shown in Figs. S12D–F.

Fig. S12G plots friction vs. scan rate. Even with high standard deviations in the friction signals of both up and down domains, the average friction signal is always higher in the up domain than the down domain. Figs. S12I and S12J show the etch depth vs. scan rate in Figs. S12A and S12C, respectively. The wear depth again decreases with scan order, not with the scan rate change.

Computational model

We follow the linear continuum model of piezoelectricity augmented with flexoelectricity (*44*). The electromechanical enthalpy density of a dielectric solid exhibiting piezoelectricity and flexoelectricity considered here, in terms of the strain $\varepsilon$, the strain gradient $\nabla\varepsilon$, and the electric field $E$, is given by

$$\Psi(\varepsilon, \nabla\varepsilon, E) = \frac{1}{2}C_{ijkl}\ \varepsilon_{ij}\ \varepsilon_{kl} - e_{ijk}\ E_i\varepsilon_{kl} - \mu_{ijkl}\ E_i\varepsilon_{jk,l} - \frac{1}{2}\kappa_{ij}\ E_iE_j +$$
$$\frac{1}{2}h_{ijklmn}\ \varepsilon_{ij,k}\varepsilon_{lm,n} \quad (S1)$$

where $C$ is the fourth-rank elasticity tensor, $e$ is the third-rank piezoelectricity tensor, $\mu$ is the fourth-rank flexoelectricity tensor, $\kappa$ is the second-rank dielectricity tensor and $h$ is the sixth-rank strain gradient elasticity tensor. The strain gradient elasticity term is required to guarantee the thermodynamic stability of the model in the presence of flexoelectricity (*48–50*). Note that following (*32*), strain $\varepsilon$, and thus the polarization $p$, are taken relative to the remanent state of the solid, $E_r$, $p_r$.

The associated free enthalpy of the system is given by

$$\Pi[\boldsymbol{u},\phi] = \int_\Omega \Psi\,d\Omega - W^{ext}, \quad (S2)$$

where $\boldsymbol{u}^*$ is the displacement, $\phi$ is the electric potential $E = -\nabla\phi$, $\Omega$ denotes the solid domain, and $W^{ext}$ is the work of external forces and charges. The Euler-Lagrange equations follow from the variational principle $(\boldsymbol{u}^*,\phi^*) = \arg\min_{\boldsymbol{u}}\max_{\phi}\Pi[\boldsymbol{u},\phi]$ as (*31*)

$$\begin{cases} (\hat{\sigma}_{ij} - \tilde{\sigma}_{ijk,k})_{,j} + b_i = 0, & \text{in } \Omega \\ D_{l,l} - q = 0, & \text{in } \Omega, \end{cases} \quad (S3)$$

where $\boldsymbol{b}$ is the external body forces per unit volume, $q$ the external free charges per unit volume, and the stress $\hat{\boldsymbol{\sigma}}$, double stress $\tilde{\boldsymbol{\sigma}}$, and electric displacement $\boldsymbol{D}$

$$\hat{\sigma}_{ij}(\boldsymbol{u}, \phi) = \left.\frac{\partial \Psi[\boldsymbol{\varepsilon}, \nabla\boldsymbol{\varepsilon}, \boldsymbol{E}]}{\partial \varepsilon_{ij}}\right|_{\substack{\nabla\varepsilon \\ E}} = C_{ijkl}\, E_{kl} - e_{ijk}\, E_i, \qquad (S4)$$

$$\tilde{\sigma}_{ijk}(\boldsymbol{u}, \phi) = \left.\frac{\partial \Psi[\boldsymbol{\varepsilon}, \nabla\boldsymbol{\varepsilon}, \boldsymbol{E}]}{\partial \varepsilon_{ij,k}}\right|_{\substack{\varepsilon \\ E}} = h_{ijklm\,n} E_{m,n} - \mu_{ijkl}\, E_i, \qquad (S5)$$

$$D_l(\boldsymbol{u}, \phi) = -\left.\frac{\partial \Psi[\boldsymbol{\varepsilon}, \nabla\boldsymbol{\varepsilon}, \boldsymbol{E}]}{\partial E_l}\right|_{\substack{\nabla\varepsilon \\ E}} = \kappa_{lm}\, E_m + e_{lij}\, E_{ij} + \mu_{ijk}\, E_{ij,k}. \qquad (S6)$$

Note that the physical stress is $\boldsymbol{\sigma} = \hat{\boldsymbol{\sigma}} - \nabla\tilde{\boldsymbol{\sigma}}$. Equations S3 are subject to appropriate boundary conditions

$$\begin{aligned} u_i &= \bar{u}_i \text{ on } \Gamma_u & t_i &= \bar{t}_i \text{ on } \Gamma_t, \\ \partial^n u_i &= \bar{v}_i \text{ on } \Gamma_v & r_i &= \bar{r}_i \text{ on } \Gamma_r, \\ \phi &= \bar{\phi} \text{ on } \Gamma_\phi & w &= \bar{w} \text{ on } \Gamma_w, \\ u_i &= \bar{u}_i \text{ on } C_u & j_i &= \bar{j}_i \text{ on } C_j, \end{aligned} \qquad (S7)$$

with the normal derivative operator $\partial^n(A) := \partial A/\partial \boldsymbol{n}$. The overline indicates a prescribed quantity on the first and second order Dirichlet and Neumann boundaries, $\Gamma_u, \Gamma_t, \Gamma_r, \Gamma_v$ with $\partial\Omega = \Gamma_t \cup \Gamma_u = \Gamma_r \cup \Gamma_v$, and $\partial\Omega$ being the boundary of the domain $\Omega$. The traction $\boldsymbol{t}(\boldsymbol{u}, \phi)$, double traction $\boldsymbol{r}(\boldsymbol{u}, \phi)$, electric charge density $w(\boldsymbol{u}, \phi)$ and edge forces $\boldsymbol{j}(\boldsymbol{u}, \phi)$ (*31, 51*) are given by

$$\begin{aligned} t_i(\boldsymbol{u}, \phi) &= \left(\hat{\sigma}_{ij} - \tilde{\sigma}_{ijk,k} - \tilde{\sigma}_{kj,l}(\delta_k - n_l n_k)\right)n_j + \tilde{\sigma}_{ijk}\, \tilde{N}_{jk} \text{ on } \partial\Omega, & (S8a) \\ r_i(\boldsymbol{u}, \phi) &= \tilde{\sigma}_{ijk}\, n_j n_k & \text{on } \partial\Omega & (S8b) \\ w(\boldsymbol{u}, \phi) &= -\hat{D}_l n_l & \text{on } \partial\Omega & (S8c) \\ j_i(\boldsymbol{u}, \phi) &= \tilde{\sigma}_{ijk}\left(m_j^L n_k^L + m_j^R n_k^R\right) & \text{on } \partial\Omega & (S8d) \end{aligned}$$

where $\tilde{N}_{ij} = -n_{i,l}(\delta_{lj} - n_l n_j) + n_{f,g}(\delta_{fg} - n_f n_g)n_i n_j$, $\partial\Omega$ is the boundary of the domain, $C_\Omega$ is the set of all edges (corners in 2D) of the boundary, and **n** is the unitary exterior normal vector. The superscripts L and R respectively refer to the first and second surfaces sharing the edge, and **m** is the conormal vector on each surface, tangent to the surface, normal to the edge and pointing outward of the surface. In 2D, **m** is simply the vector tangent to the side and pointing outward on the corner. The domain boundary is assumed to be composed of smooth surfaces (curves in 2D) and joints at sharp boundary edges (corners in 2D). $C_j$ denotes the union of the boundary edges that are shared by two surfaces with first order Neumann conditions, i.e. the edges of $\Gamma_t$, where a line (punctual in 2D) force $\bar{j}_i$ is set. $C_u$ denotes the union of all other edges, i.e. those shared by at least one Dirichlet surface, where the value of $u_i$ is assumed to be that of the adjacent Dirichlet surface, i.e., $u_i = \underline{u}_i$ on $C_u \subset \bar{\Gamma}_u$.

The weak form of the problem is obtained from the first-order stationary conditions

$$\delta\Pi[\mathbf{u}, \phi; \delta\mathbf{u}, \delta\phi] = 0 \qquad \forall \delta\mathbf{u} \in H^2(\Omega), \quad \forall \delta\phi \in H^1(\Omega). \qquad (S9)$$

In particular, for the 2D axisymmetric problem with rotationally symmetric structures and axisymmetric material tensors under axisymmetric loading,

$$\delta\Pi^\Omega[\mathbf{u}, \phi; \delta\mathbf{u}, \delta\phi] = 2\pi \int_\Omega \left(\hat{\sigma}_{ij}\delta\varepsilon_{ij} + \tilde{\sigma}_{ijk}\delta\varepsilon_{ij,k} - \hat{D}_l \delta E_l\right) r\, dr\, dz \qquad (S10)$$

where the strain, strain gradient and electric field in axisymmetric notation $(r, \phi, z)$ are defined element-wise as (*52, 53*)

$$\varepsilon_{rr} = \frac{\partial u_r}{\partial r}, \qquad \varepsilon_{rz} = \frac{1}{2}\left(\frac{\partial u_r}{\partial z} + \frac{\partial u_z}{\partial r}\right),$$

$$\varepsilon_{zz} = \frac{\partial u_z}{\partial z}, \qquad \varepsilon_{\phi\phi} = \frac{u_r}{r}, \qquad (S11)$$

$$\varepsilon_{rr,r} = \frac{\partial^2 u_r}{\partial r^2},$$

$$\varepsilon_{\phi\phi,r} = \frac{1}{r}\frac{\partial u_r}{\partial r} - \frac{u_r}{2r^2},$$

$$\varepsilon_{r\phi,\phi} = \varepsilon_{\phi r,\phi} = \frac{1}{r}\frac{\partial u_r}{\partial r} - \frac{3u_r}{4r^2},$$

$$\varepsilon_{\phi\phi,z} = \frac{1}{r}\frac{\partial u_r}{\partial z},$$

$$\varepsilon_{rr,z} = \frac{\partial^2 u_r}{\partial r \partial z},$$

$$\varepsilon_{rz,r} = \varepsilon_{zr,r} = \frac{1}{2}\left(\frac{\partial^2 u_r}{\partial r \partial z} + \frac{\partial^2 u_z}{\partial r^2}\right),$$

$$\varepsilon_{zz,r} = \frac{\partial^2 u_z}{\partial r \partial z},$$

$$\varepsilon_{z\phi,\phi} = \varepsilon_{\phi z,\phi} = \frac{1}{2r}\left(\frac{\partial u_r}{\partial z} + \frac{\partial u_z}{\partial r}\right),$$

$$\varepsilon_{rz,z} = \varepsilon_{zr,z} = \frac{1}{2}\left(\frac{\partial^2 u_z}{\partial r \partial z} + \frac{\partial^2 u_z}{\partial z^2}\right),$$

$$\varepsilon_{zz,z} = \frac{\partial^2 u_z}{\partial z^2}, \qquad (S12)$$

$$E_r = -\frac{\partial \phi}{\partial r}, \qquad\qquad E_z = -\frac{\partial \phi}{\partial z}. \qquad (S13)$$

In addition, the following conditions ensure a well-posed saddle point problem:

$$\delta_u^2 \Pi[\mathbf{u}, \phi; \delta\mathbf{u}] > 0, \quad \delta_\phi^2 \Pi[\mathbf{u}, \phi; \delta\phi] < 0, \quad \forall \delta\mathbf{u}, \delta\phi. \quad (S14)$$

Finally, the weak form of the problem is given by

$$\text{Find } (\mathbf{u}, \phi) \in H^2(\Omega) \otimes H^1(\Omega) \text{ such that } \delta\Pi[\mathbf{u}, \phi; \delta\mathbf{u}, \delta\phi] = 0,$$
$$\forall (\delta\mathbf{u}, \delta\phi) \in H^2(\Omega) \otimes H^1(\Omega). \quad (S15)$$

We expand the continuum displacement and electric potential fields as

$$u_i(x) = \sum_{a=1}^{N} B^a(x) u_i^a, \qquad \phi(x) = \sum_{a=1}^{N} B^a(x) \phi^a, \quad (S16)$$

where $\mathbf{u}^a$ and $\phi^a$ are the values of the displacement and electric potential of the a-th basis function, respectively. The constitutive equations of the model (S4), (S5) and (S6) require at least $C^1$ continuity. To achieve the required continuity, we use b-splines basis functions. More precisely, we consider piecewise polynomial functions with $C^{p-1}$ continuity, with $p > 2$ being the degree of approximation. The univariate b-splines are $\{B_i^p\}_{i=0}^{n_\xi - 1}$ and they are defined in a parametric space $\xi \in [0, n_\xi]$ with the following recursive formula:

$$B_i^0(\xi) = \begin{cases} 1 & \xi_i \leq \xi < \xi_{i+1} \\ 0 & \text{otherwise} \end{cases}; \quad B_i^k = \frac{\xi - \xi_i}{\xi_{i+k} - \xi_i} B_i^{k-1}(\xi) + \frac{\xi_{i+k+1} - \xi}{\xi_{i+k+1} - \xi_{i+1}} B_{i+1}^{k-1}(\xi); \quad (S17)$$

for $k = 1, \ldots, p$ and $i = 0, \ldots, n_\xi + p - k - 1$, where $\{\xi_i\}_{i=0}^{n_\xi - 1}$ are the so-called knot points. B-splines are defined in a multivariate space by the tensor product of univariate ones, i.e.,

$$B_{[i_{xi}, i_\eta]}^p([\xi, \eta]) = B_{i_\xi}^p(\xi) B_{i_\eta}^p(\eta); \quad i_\xi = 0, \ldots, n_\xi - 1; \quad i_\eta = 0, \ldots, n_\eta - 1. \quad (S18)$$

Contact model

The Signorini-Hertz-Moreau model (*54, 55*) for a pair of points in contact $(x; x_0)$ as

$$g_n \leq 0, \quad \sigma_n = \mathbf{n} \cdot \boldsymbol{\sigma} \cdot \mathbf{n}, \quad g_n \sigma_n = 0 \text{ on } \Gamma_c \quad \text{(S19)}$$

where **n** is the normal vector to the contact boundary $\Gamma_c$ and $g_n$ is the gap function described as

$$g_n = \mathbf{u} \cdot \mathbf{n} + (\mathbf{x} - \mathbf{x}_0) \cdot \mathbf{n} \quad \text{(S20)}$$

with $x$ being a point on the flat surface of contact and $x_0$ the corresponding point of the rigid cone that is coming into contact. To enforce the non-penetration constraint, we add an extra surface energy term penalizing deviation from the conditions in Eq. (S19), yielding the final total enthalpy as

$$\overline{\Pi} = \Pi + \int_{\Gamma_c} \frac{1}{2} \beta \langle g_n \rangle^2 d\Gamma ; \quad \text{(S21)}$$

where $\beta$ is a penalty parameter and $\langle A \rangle$ is the Macaulay bracket which returns the argument itself if the argument is positive and 0 otherwise. The rotational symmetry of the conical tip sample allows the use of the two-dimensional axisymmetric model. The contact force $F$ is then obtained as

$$F = 2\pi \int_0^{R_c} \beta \langle g_n \rangle r dr . \quad \text{(S22)}$$

We verify the model against the classical conical contact model in (*56*), for the case of an elastic solid with vanishing piezoelectricity $\mathbf{e}$, flexoelectricity $\boldsymbol{\mu}$, dielectricity $\boldsymbol{\kappa}$, and strain gradient elasticity $\mathbf{h}$ (Eq. (S1)). We perform simulations for different vertical positions of the tip and compute the magnitude of the contact force and the contact radius. The relation between the indentation depth $d$ and the contact radius $R_c$ is given by

$$d(R_c) = \frac{\pi}{2} R_c \tan \theta, \quad \text{(S23)}$$

where $\theta = 42.5°$ corresponds to the nominal value provided by the manufacturer. The geometrical parameters of the simulation are shown in Fig. S14. The relation between the applied force $F$ and the contact radius is given by

$$F = \frac{\pi R_c^2}{2} E^* \tan\theta, \qquad (S24)$$

where $E^* = \frac{E}{1-v^2}$, with $E = 164$ GPa and $v = 0.3$ in this simulation. We have considered a rectangular elastic solid of length $L = 100$ nm and height $H = 50$ nm in Fig. S14. We find perfect agreement between the theoretical model and the computational one as shown in Fig. S15.

Flexoelectric contact simulations.

Fig. S18 shows plots of the indentation depth and contact radius as functions of applied force. Apparently, in the absence of flexoelectricity, both quantities coincide for upward and downward polarized domains, as expected. Flexoelectricity breaks this symmetry and a noticeable difference in the response is found for up and down domains. This is further illustrated in Fig. S17, where the flexoelectrically induced polarization and piezoelectric polarization are plotted separately, for the case of upward and downward polarized domains. While the flexoelectric polarization is largely independent of the direction of ferroelectric polarization, the piezoelectric polarization has opposite signs for up and down domains. Hence, their combination is the origin of the observed asymmetric response. As indicated by previous studies on electromechanical response and fracture in dielectrics (*26, 57*), the asymmetry resulting from the interaction of flexoelectricity and piezoelectricity is observed only in a range where these two mechanisms compete in magnitude. If either of them clearly dominates the other, i.e. for very small or very large flexoelectricity with respect to piezoelectricity, the asymmetry disappears, as the response of each mechanism is not asymmetric by itself. Within the range where flexoelectric and piezoelectric polarizations are comparable in magnitude, the observed difference in the response for up and down domains depends on the magnitude of flexoelectric parameters, being larger for stronger flexoelectricity (Fig. S19).

I–V curve measurements

To select a probe with negligible conductivity for the mechanism study of asymmetric tribology, we performed I–V curve measurements on a highly ordered pyrolytic graphite sample using six different types of probes: a Pt/Ir-coated Si probe (EFM, *NanoWorld*), diamond coated Si probes (DT-NCHR and CDT-NCHR, *NANOSENSORS*), a diamond-like-carbon (DLC)-coated probe (HQ:NSC16/HARD/Al BS, *MikroMasch*) and single crystalline diamond probes (NM-TC, *Adama Innovations* and D300, *SCD Probes*). EFM, DT-NCHR, CDT-NCHR and NM-TC show conductive behavior as can be seen in Figs. S20A and S20B. DLC coated probe and D300 probe show almost negligible current, but a closer look at I–V curves in Fig. S20C shows that D300 has the most negligible conductivity. D300 shows this insulating characteristic because the single crystalline diamond is attached to the cantilever with non-conductive glue. The probe selection for the asymmetric etching of PPLN is summarized with more detail in Table. S1. We successfully demonstrated asymmetric etching with five diamond probes, but not with the Pt/Ir-coated Si probe.

Asymmetric friction and wear in stoichiometric LiNbO$_3$

A third possible mechanism is that the asymmetry is defect mediated, at the origin of the asymmetric tribology response. Asymmetric friction can originate from the defect concentration difference between up and down domains as in p-n junctions, which show varying friction because of charge depletion and accumulation (*58*). The congruent PPLN used in this study is known to have relatively high densities of defects (*59*). To evaluate their possible contribution, we investigated the etching effects in stoichiometric LiNbO$_3$, where the defect density is lower than in the congruent composition. Figs. S21A–C show height, PFM phase and amplitude before milling with naturally formed up and down domains in the sample scan region. Height and friction differences are observed between up and down domains during the milling scans shown in Figs. S21D (topography) and S21E (friction). During the milling scans, the friction is higher in up domains than in down domains. As a result of 10 milling scans at a loading force of 5 μN and a scan rate of 4.88 Hz, we also observe higher topography in the down domain (Fig. S21F), showing no significant difference between PPLN and stoichiometric LiNbO$_3$. Stable polarization configuration after milling scans is shown in the PFM phase and amplitude images in Figs. S21G and S21H. Therefore, we conclude that the presence of defects is not responsible for the observed asymmetry.

Tomographical studies of LiNbO$_3$ nano pillars

We note that up domains are visible in the core of the pillars after prolonged etching of around 60 milling scans (see Supplementary Fig. 23E). This feature observed in LiNbO$_3$ nanodomain structures is a product of the electrostatic boundary conditions at the tip of the growing domain during initial switching, resulting in incomplete polarization reversal through the single crystal (*60-62*). This core domain places a limit on the pillar height which can be achieved by nanodomain patterning. As seen in Fig. 3E, while the pillars are still fully capped by a down domain (first 30 millings), their height increases with milling in a linear fashion, similar to what we observed for the wide periodic domains fully penetrating the PPLN single crystal. However, as this cap is gradually worn through, and the up polarized core is revealed, the wear rate of the pillar increases, and becomes comparable to that of the up-oriented background state in Supplementary Fig. 23E.

Ferroelectric polarization-derived nanostructuring of LiNbO$_3$ using wear asymmetry

We found that our polarization-determined structuring could be demonstrated in reduced dimensions such as 100 nm of LiNbO$_3$ thin film (*NanoLN*). To allow easily controllable domain patterning, we used a thinner LiNbO$_3$ fabricated by the ion slicing method. The pristine domain configuration is upward. Fig. S24 shows that facile fabrication without any chemicals or photomask is realized after the artificial decoration of the thin film with the text "FERRO" with down domains, and "LITHO" with up domains before etching. We used a Pt-coated Si probe (HQ:DPER-XSC11, *MikroMasch*) to switch the domain, and then changed to single-crystalline diamond probe (NM-TC, *Adama Innovations*, Lot number: 009-013) to prevent tip contamination. Multiple mechanical milling scans using a diamond probe create a height difference between up and down domains. As shown in Fig. S24D, the friction in the up domain is still higher than down domain during the milling scan at the loading force of 12.5 μN, with a friction ratio close to the simulation values in Fig. 2G. Fig. 3F-I shows the nanostructures shown in Fig. S24 with a pristine background region, which clearly indicates the height difference between etched up and down domains and pristine surface. Fig. S27 shows the 3D fabrication procedure for ferroelectric LiNbO$_3$ by repeated switching and milling. The loading force is gradually increased from 5 to 25 μN to maintain the wear efficiency at a scan speed of 4.88 Hz. Fig. S28 shows the evolution of the topography in the 3D nanostructuring in Fig. S27.

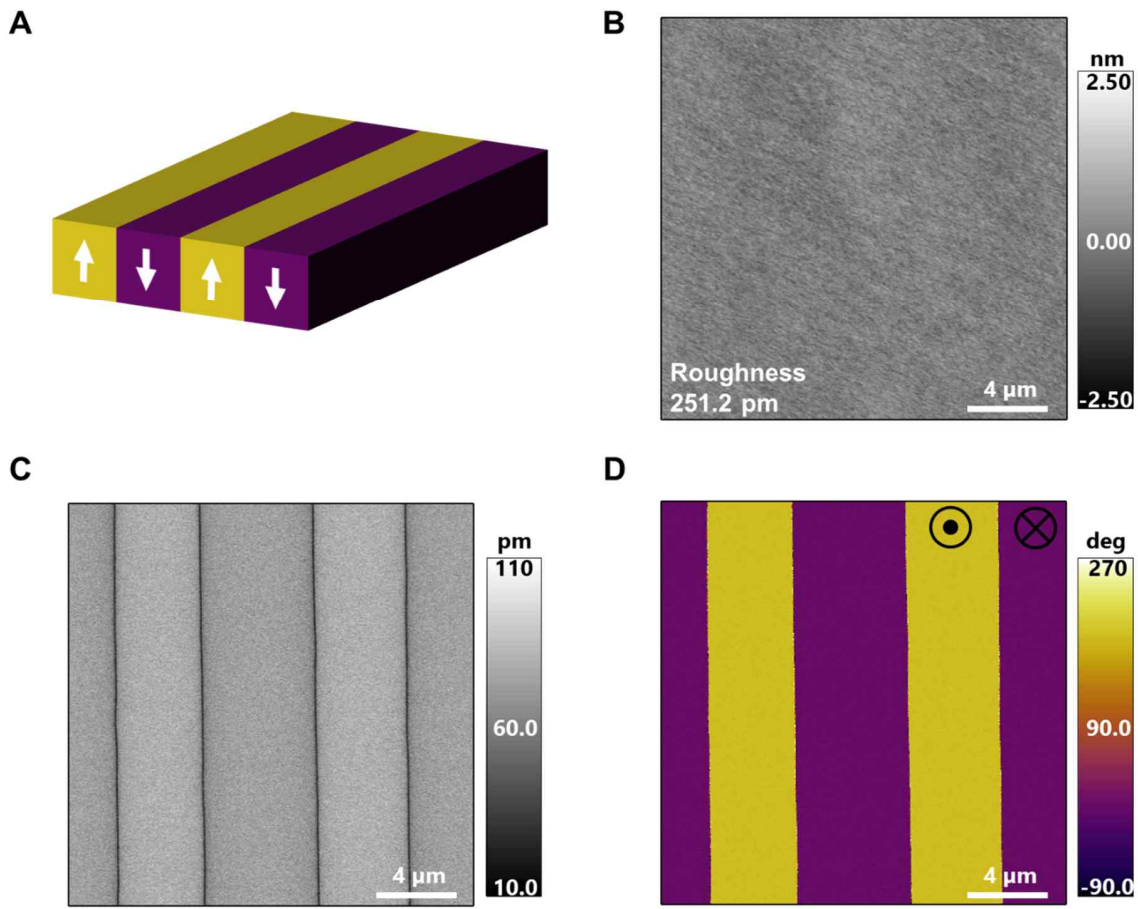

**Fig. S1. Description of pristine single crystal periodically poled LiNbO₃ (PPLN).**
**(A)** Schematic of pristine z-cut PPLN, which has uniaxial out-of-plane ferroelectric domains (up and down polarity), **(B)** height, **(C)** PFM amplitude and **(D)** PFM phase of pristine PPLN. The height shows a flat morphology of the polished surface of the crystal, and PFM amplitude and phase show the stable alternative orientation of polarization, visualized by off-resonance single frequency PFM at 17 kHz. The domain width is around 5 μm with periodic stripes.

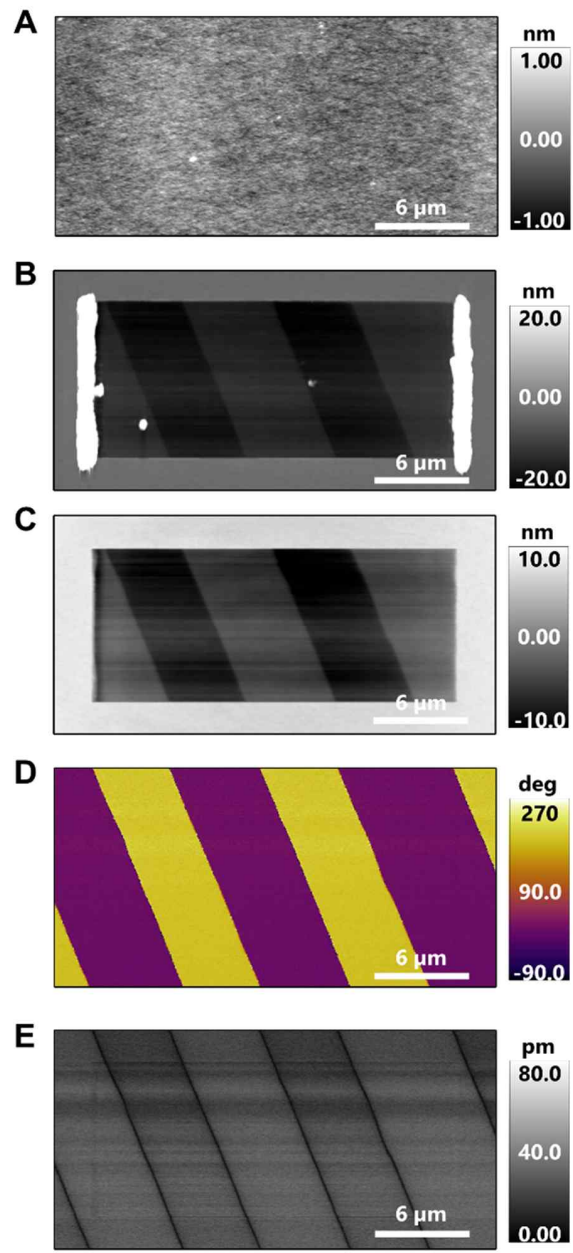

**Fig. S2. Self-cleaning of wear debris by AFM tip after milling scans.** **(A)** Height of pristine PPLN surface, **(B)** Height after continuous 40 milling scans at 8 μN. **(C)** Height after surface cleaning by AFM tip with 10 continuous contact scans at much lower loading force than that of milling. **(D)** PFM phase and **(E)** PFM amplitude after cleaning.

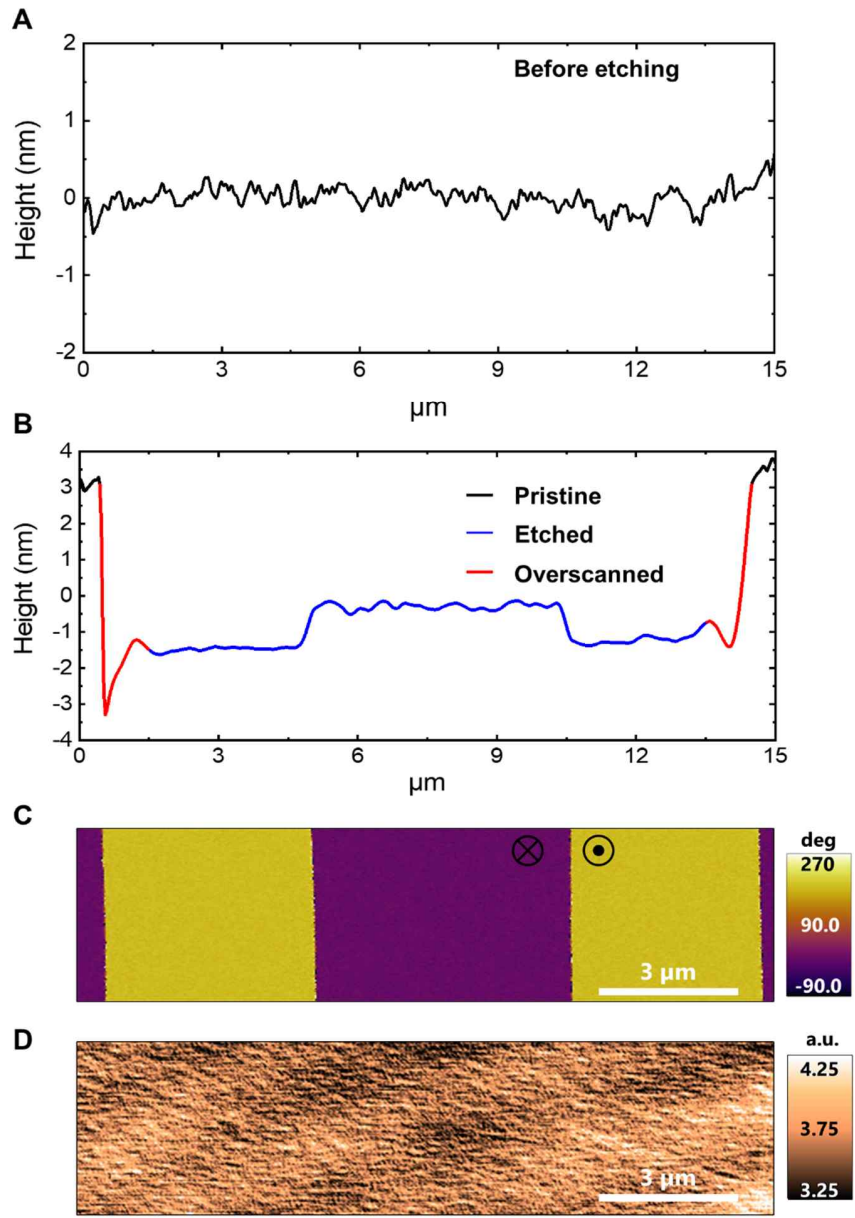

**Fig. S3. Topography before and after mechanical etching in Fig. 1 with the pristine PFM and friction images.** **(A)** Line profile along the direction perpendicular to the domain boundary before mechanical etching scans. **(B)** Line profiles after 50 milling scans: pristine (black), etched in the set region (blue) and etched in the overscanned region (red). In the overscanned region, the height is non-uniformly etched because of the distortion correction algorithm during the contact scanning. **(C)** PFM phase and **(D)** friction at approximately 300 nN before milling scans. The horizontal x-axis scales in the PFM and friction images correspond to the x-axis scales in Figs. S3A and S3B.

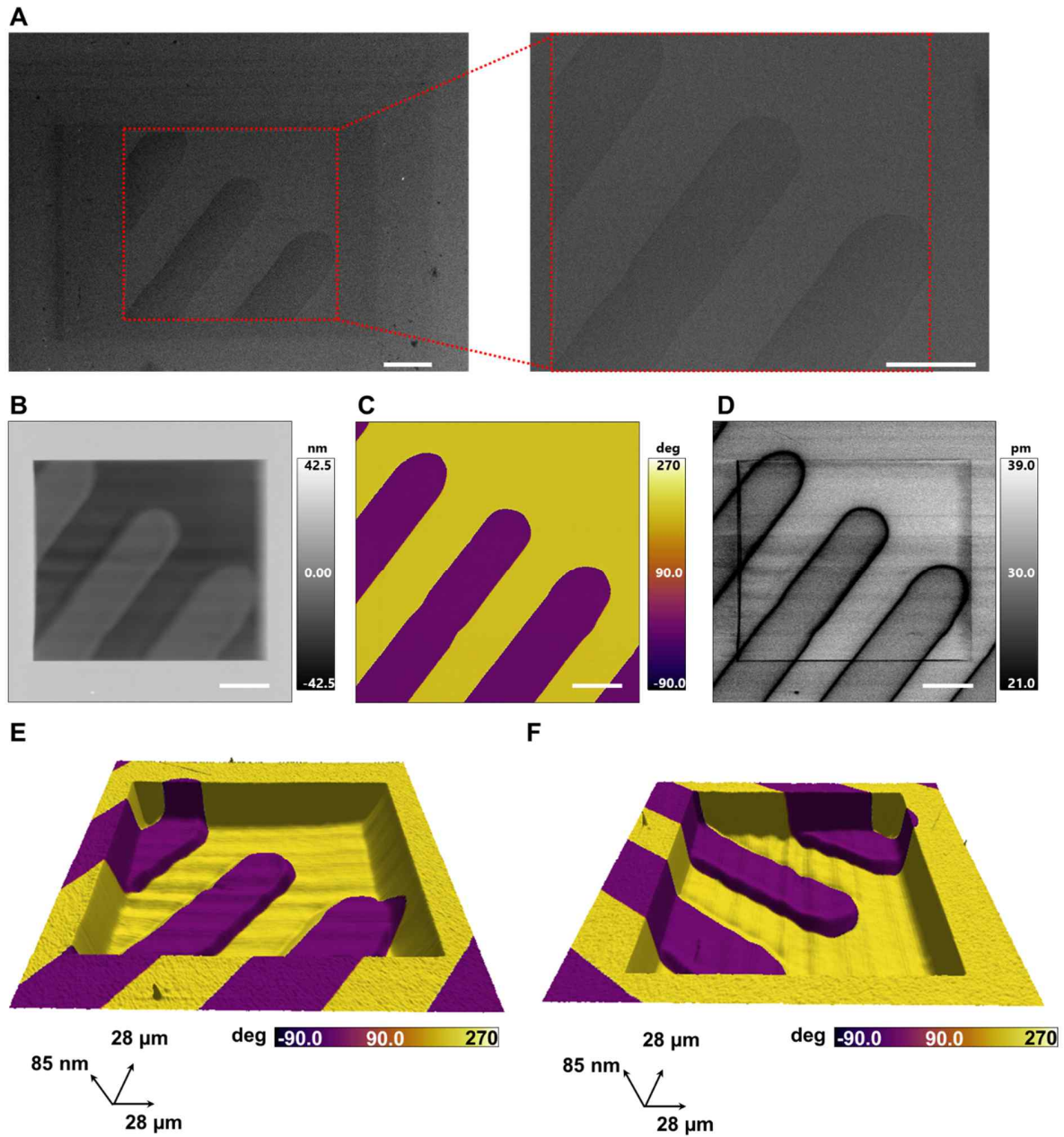

**Fig. S4. Cross-validation of wear asymmetry using SEM.** **(A)** SEM images, **(B)** Height, **(C)** PFM phase and **(D)** PFM amplitude after 50 milling scans. **(E)** 3D surface plots with color scale indicating PFM phase. SEM imaging was performed 15 days after milling scans and PFM imaging. Comparing SEM images with PFM data, the SEM contrast originates from the height difference between up and down domains. Scale bars in **(A–D)** are 5 μm.

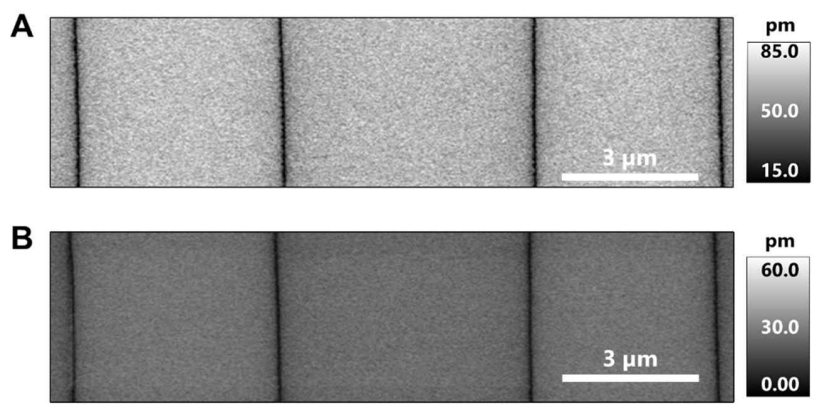

**Fig. S5. PFM amplitude before and after the mechanical etching shown in Fig. 1.** PFM amplitude **(A)** before and **(B)** after mechanical milling scans. The amplitude scale changes because of the wear debris attachment to the tip during the scans.

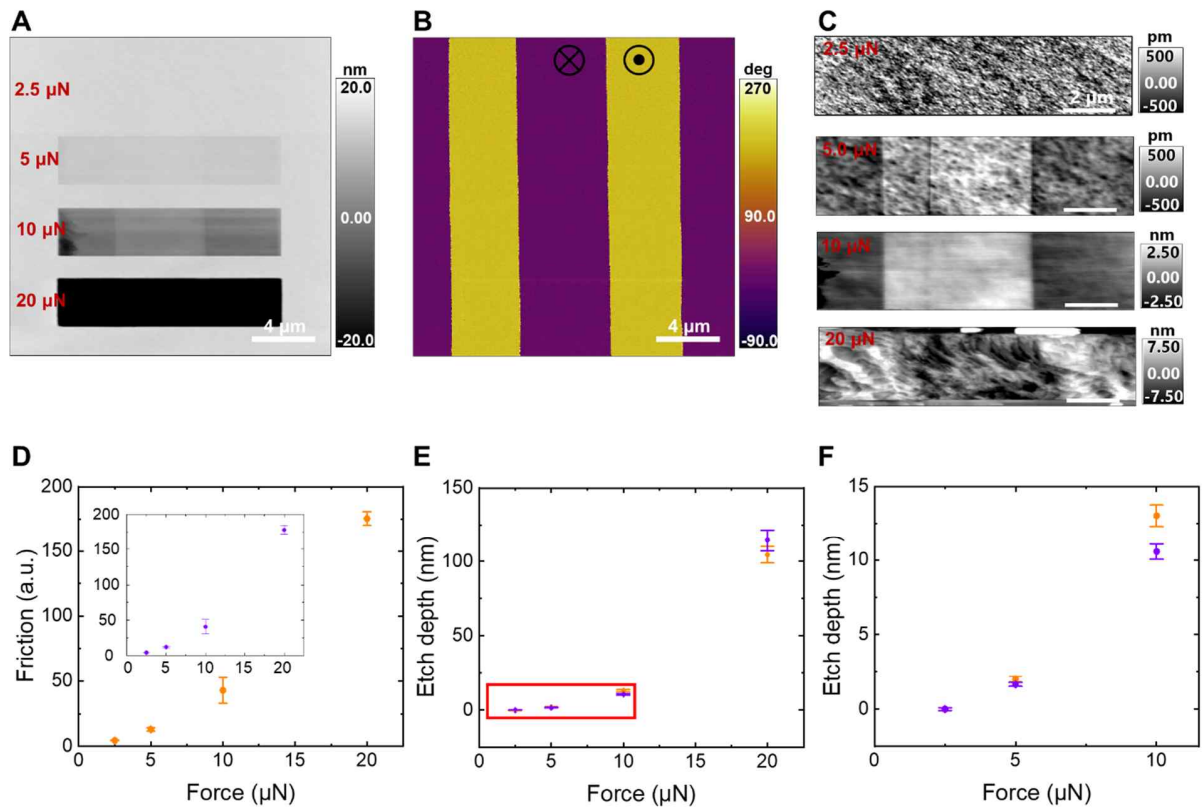

**Fig. S6**. **Contact force optimization of asymmetric friction and wear.** **(A)** Height image after ten milling scans with increasing loading forces in four different regions. **(B)** PFM phase image after ten milling scans. **(C)** Height in each region with different loading forces from 2.5 to 20 μN **(D)** Friction in up domains as a function of loading force in each region. Inset shows friction vs. loading force in down domains. **(E)** Etch depth as a function of loading force in each region. **(F)** Etch depth as a function of loading force except for the failure case.

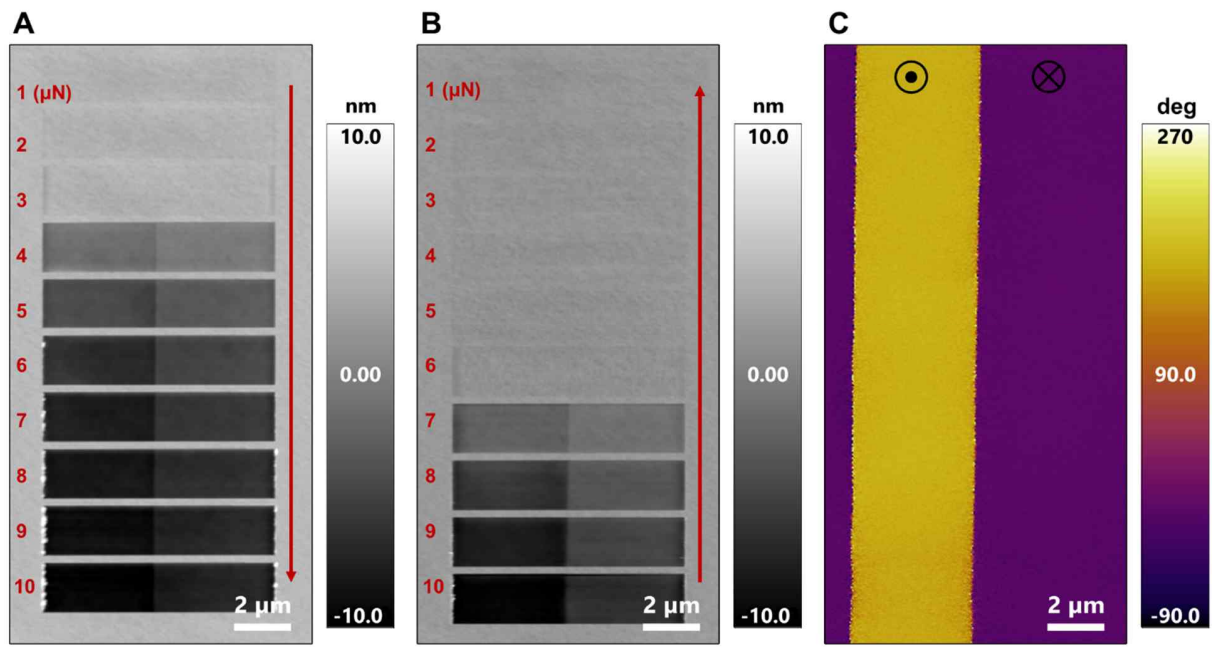

**Fig. S7. Effect of increasing and decreasing loading force on asymmetric tribology using the same probe.** Arrows indicate the order of milling scans and numbers in red font show the loading forces applied in each region. Four milling scans were conducted in each region from a loading force of 1 µN to 10 µN. **(A)** Height after milling scans with increasing loading force from 1 µN to 10 µN. **(B)** Height after milling scans with decreasing loading force from 10 µN to 1 µN. Milling scans conducted in (B) were conducted after milling scans in Fig. S7A using the same probe. **(C)** PFM phase before mechanical milling scans in the same region of Fig. S7B.

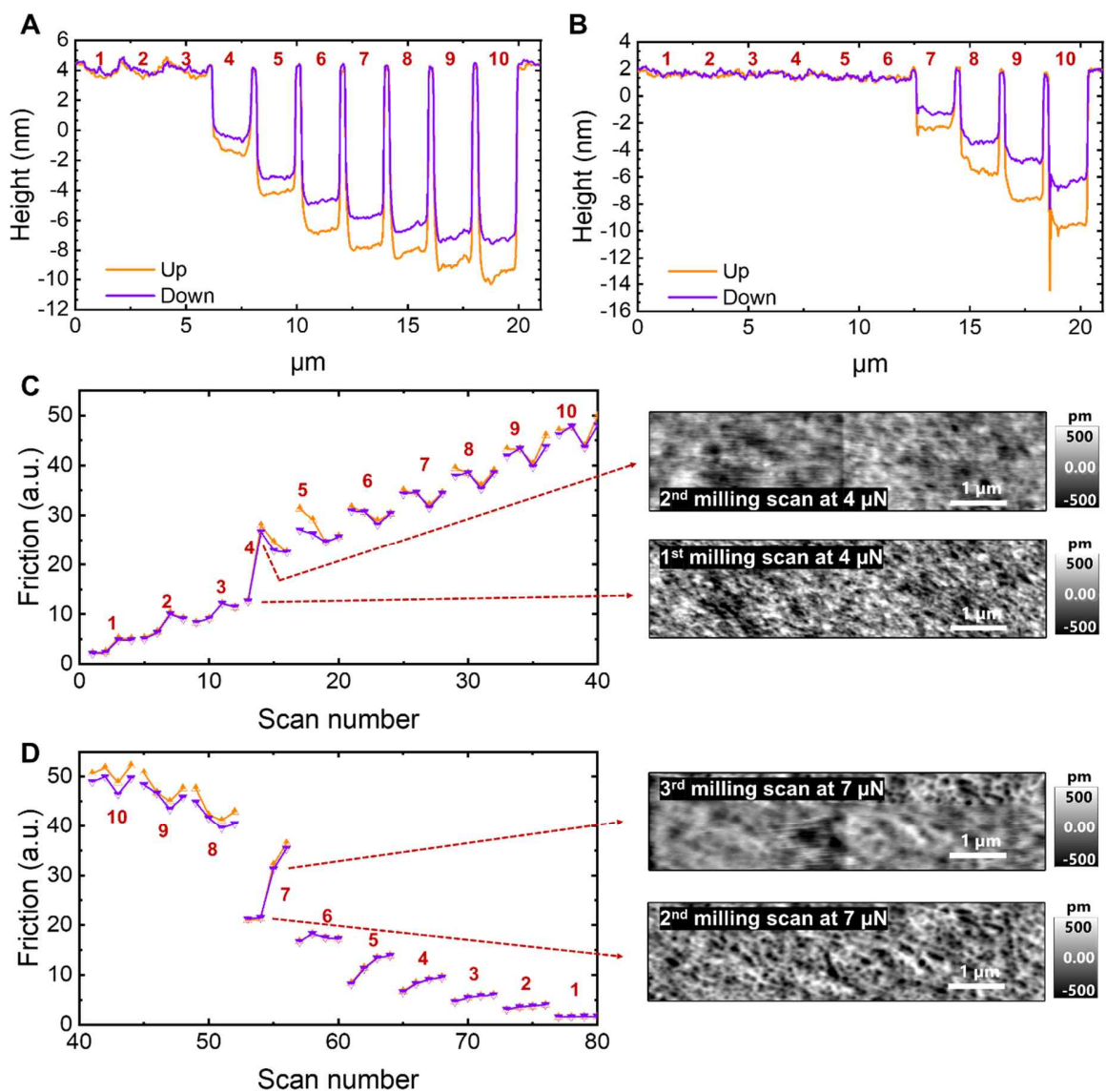

**Fig. S8. Friction and wear trend from the experiment in Fig. S7.** (**A** and **B**) Etch depth along with the up (orange) and down (purple) domains vertically in Figs. S7A and S7B, respectively. The numbers in red font indicate the loading force at each region. Four milling scans were performed in each region. (**C**) Friction vs. scan numbers for increasing loading force in Fig. S7A. Height during the 1st and 2nd milling scan at 4 µN, respectively. (**D**) Friction vs. scan numbers with decreasing loading force in Fig. S7B. Height during the 2nd and 3rd milling scan at 7 µN, respectively. Wear asymmetry is observed with the jump of friction signals seen in Figs. S8C and S8D.

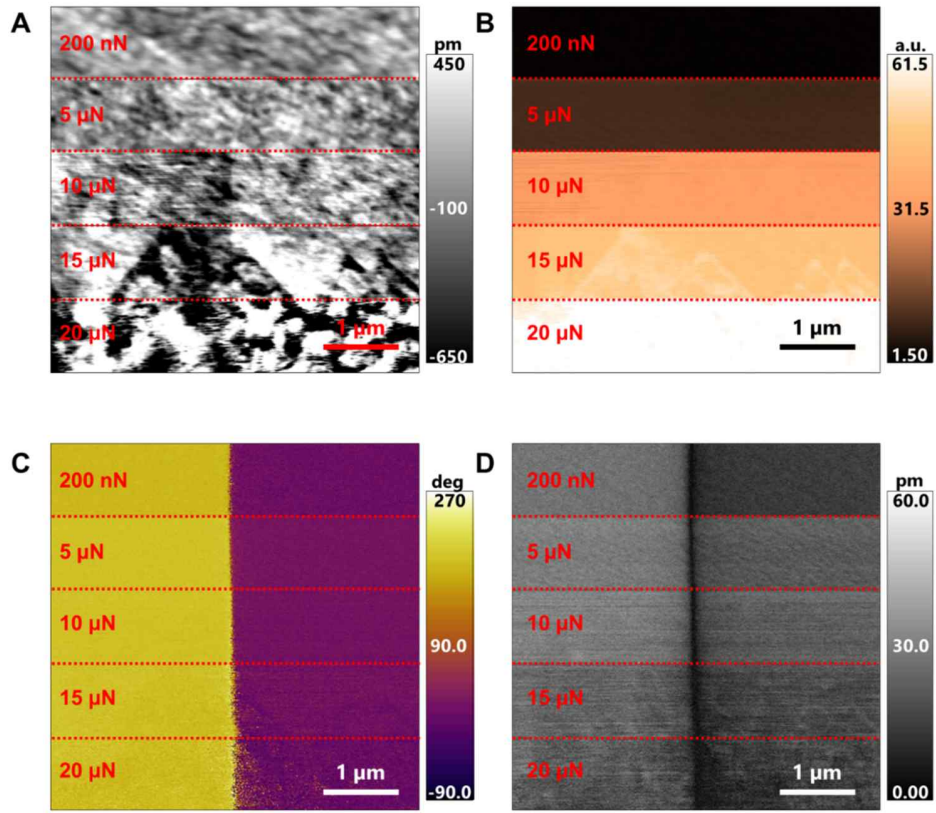

**Fig. S9. Simultaneous PFM imaging of PPLN with increasing loading force from 200 nN to 20 µN showing no transient domain switching during the application of high strain gradient.** **(A)** Topography, **(B)** friction, **(C)** PFM phase, and **(D)** PFM amplitude during the PFM scanning with increasing loading force. At high loading forces such as 15 µN and 20 µN, flexoelectric switching is not observed even with materials fracture during the scan.

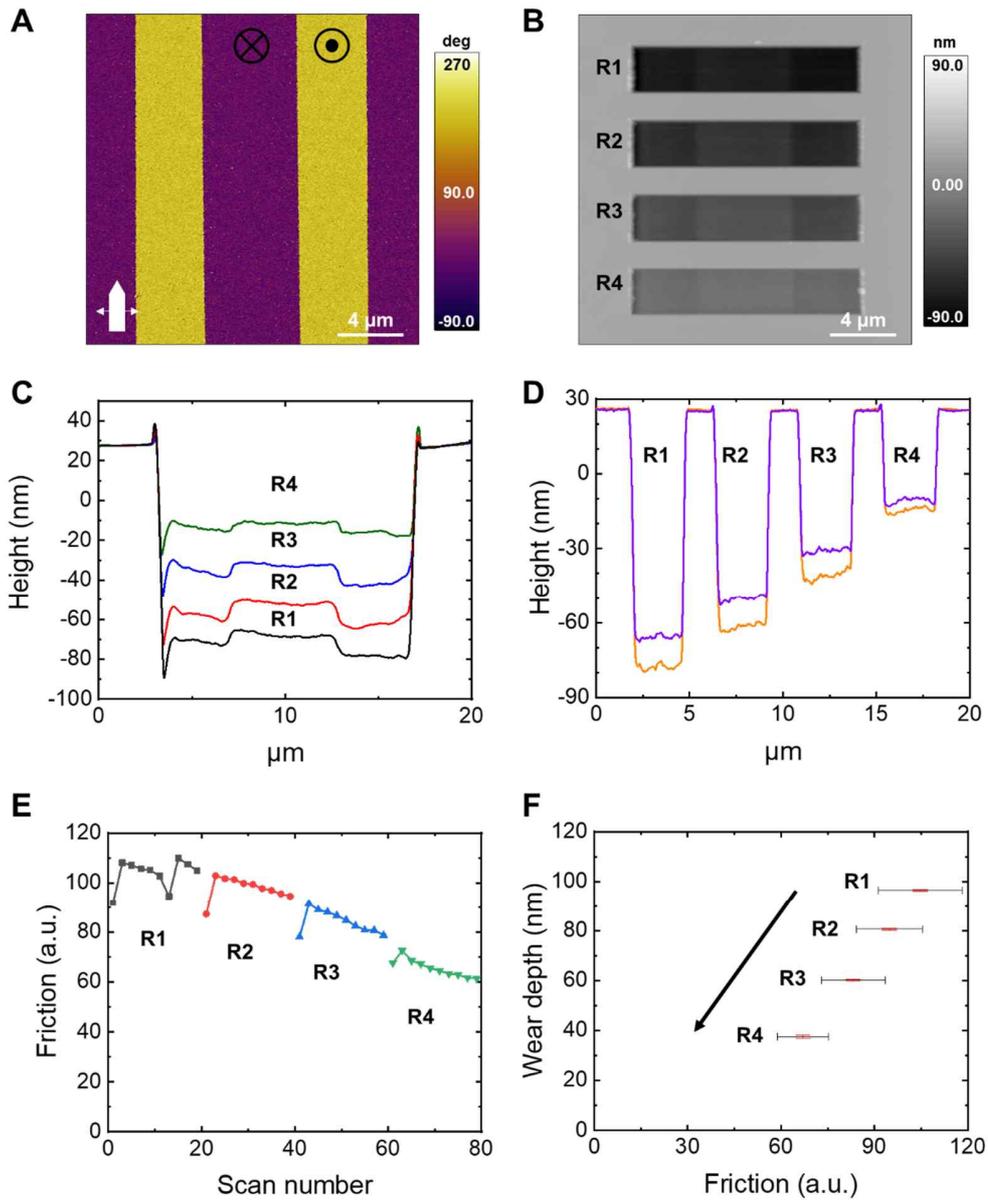

**Fig. S10. Friction and wear degradation with continuous scans.** **(A)** PFM phase image before milling scans, **(B)** height image after milling on four different regions from region 1 (R1) to region 4 (R4) with the same polarity. A loading force of 5 μN was applied at a scan rate of 1.95 Hz in each scan, and 10 milling scans were performed in each region. **(C)** Horizontal line profile along the etched regions. **(D)** Vertical line profiles along the etched height indicate the wear rate difference of up (orange) and down domains (purple). **(E)** Friction signal degradation from R1 to R4. **(F)** Correlation of average friction and wear with standard deviations. All scans are measured at 20°C and 20% relative humidity.

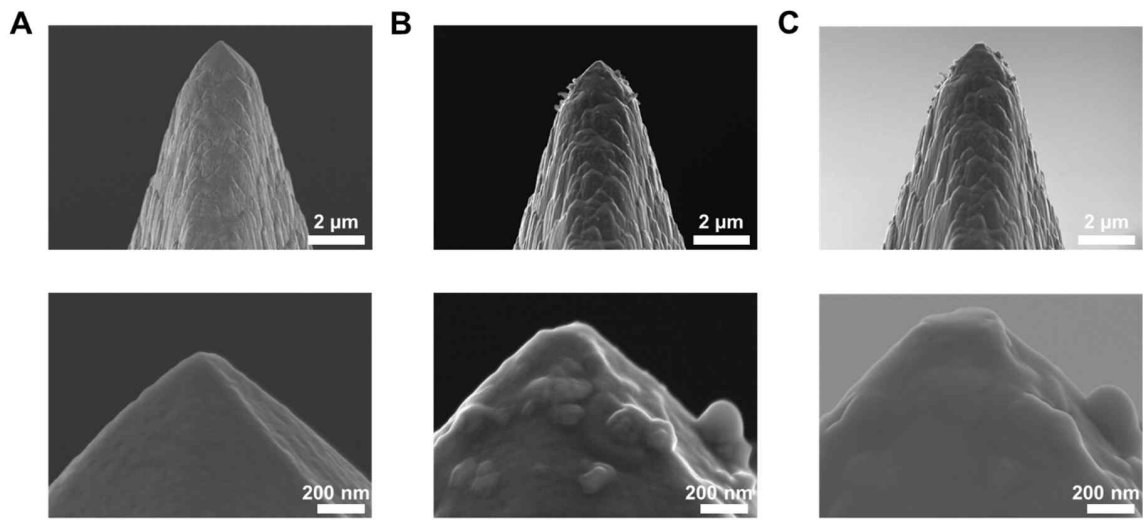

**Fig. S11. SEM images of the diamond probe before and after milling. (A)** Pristine, **(B)** after 10 millings, **(C)** after 50 millings at a loading force of 10 μN and a scan rate of 4.88 Hz on 20 μm × 20 μm of PPLN surface. The attachment of wear debris during the continuous millings is evident. The end of the probe has a dull shape covered by debris from the sharp cone shape. The debris coverage on the diamond probe is correlated with the degradation of the friction and etch depth during the continuous milling.

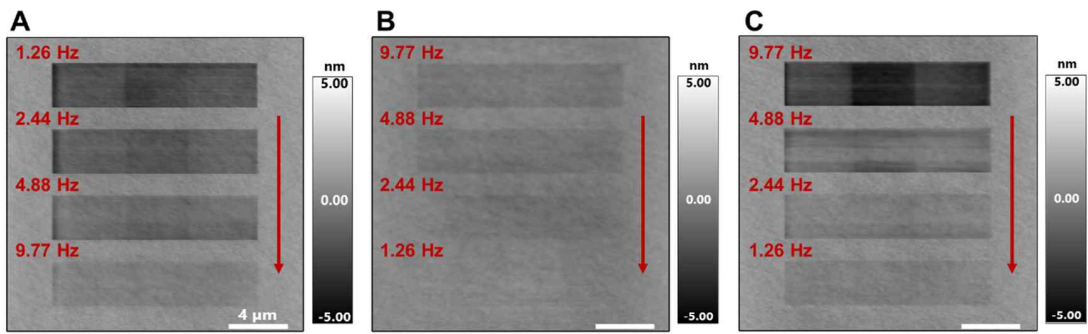
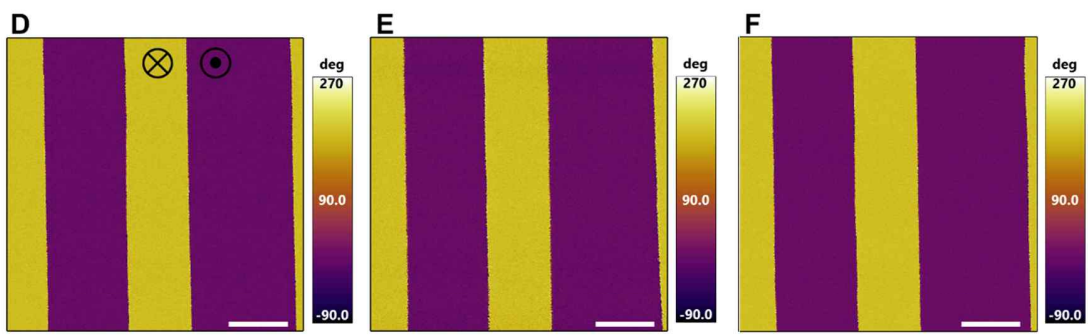
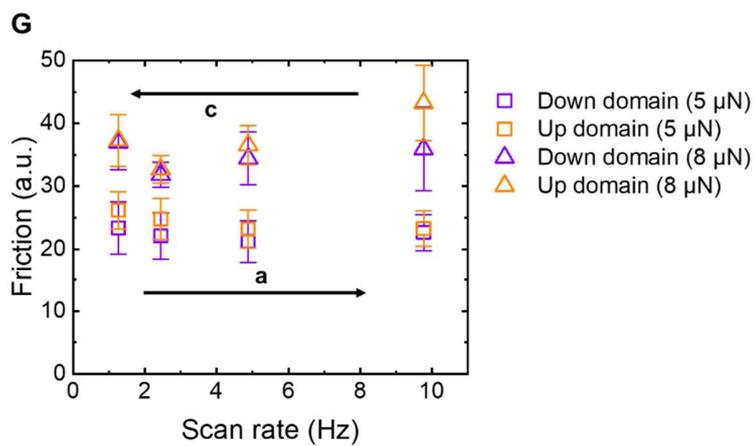
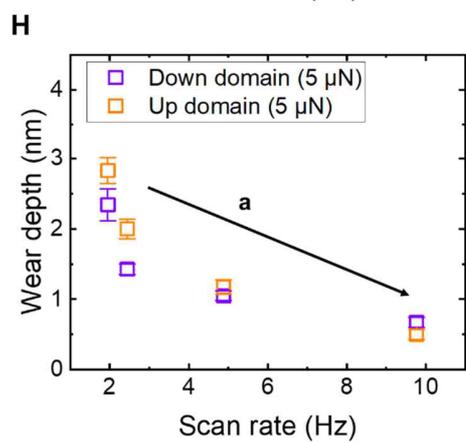
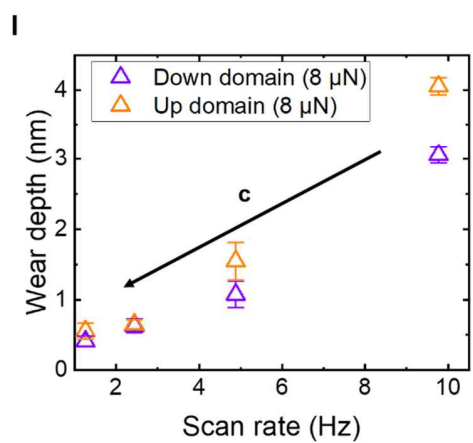

**Fig. S12. Effect of scan rate on asymmetric tribology.** Red arrows indicate the order of milling scans and numbers in red font show the scan rate applied in each region. **(A)** Height image after ten milling scans with four different scan rates at loading force of 5 μN. **(B)** Height image after ten milling scans with four different scan rates at loading force of 5 μN after milling scans in Fig. S12A using the same probe. **(C)** Height image after ten milling scans with four different scan rates at loading force of 8 μN after milling scans in Fig. S12B using the same probe. Red arrows indicate the scan order. **(D–F)** PFM phase images of Fig. S12A–C, respectively. **(G)** Friction vs. scan rate in Fig. S12A and Fig. S12C. **(H)** Wear depth depending on scan rate in Fig. S12A. **(I)** Wear depth depending on scan rate in Fig. S12C. Scan order is indicated with arrows in Figs. S12G–I.

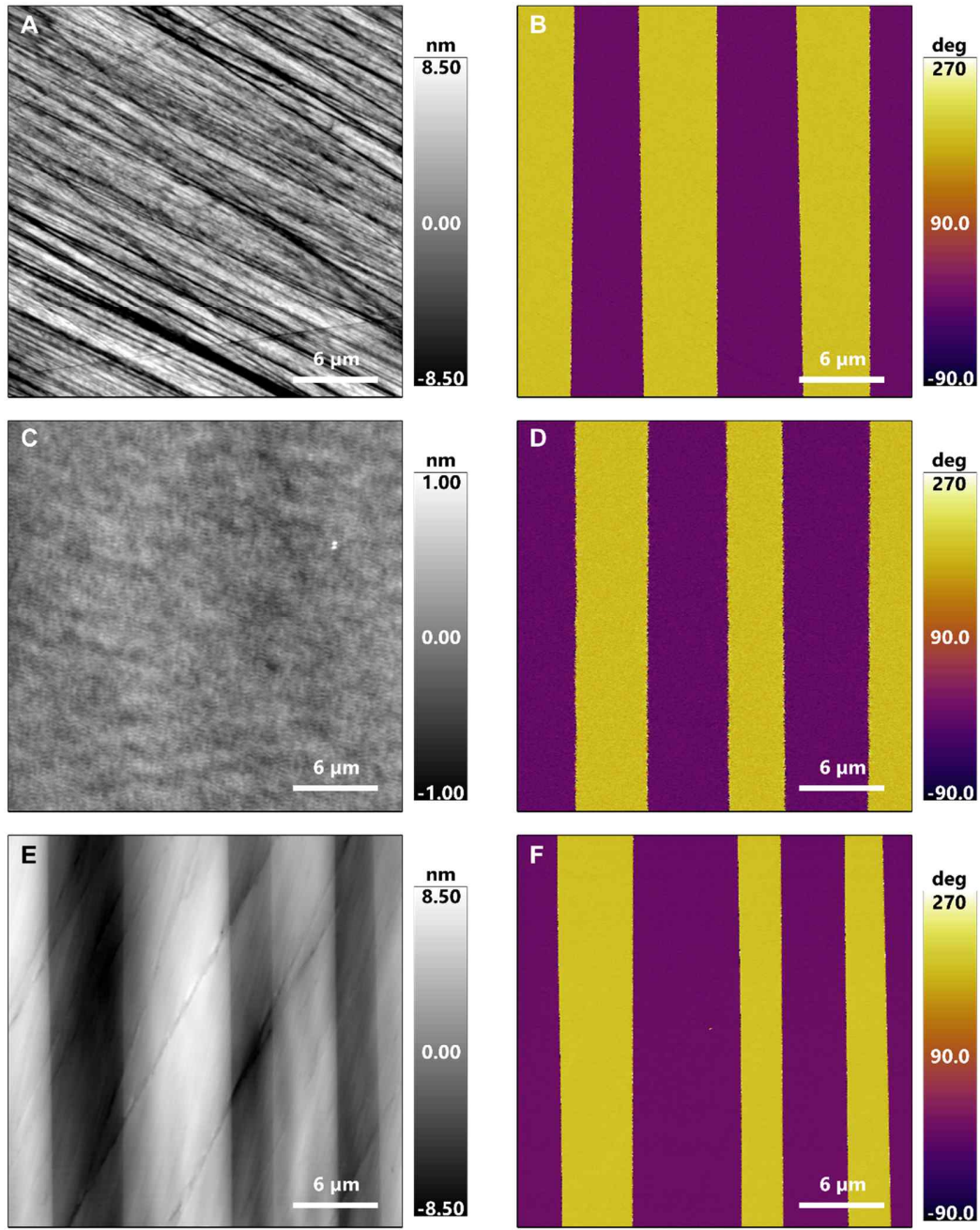

**Fig. S13. Control experiments of scalable large area milling of PPLN. (A)** Height and **(B)** PFM phase after mechanical grinding using diamond lapping films. **(C)** Height and **(D)** PFM phase after dipping in colloidal silica solution for 12 h. **(E)** Height and **(F)** PFM phase after polishing using silica nanoparticles for 3 min.

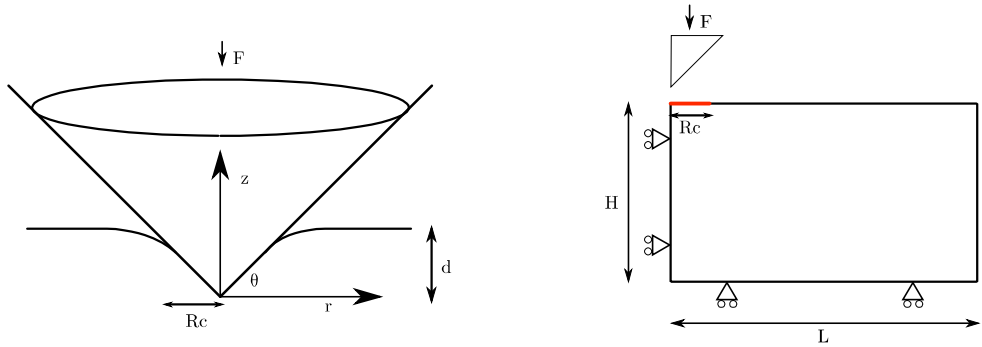

**Fig. S14. Geometrical parameters of the conical indenter (left) and axisymmetric problem statement with mechanical boundary conditions (right).** A rectangle of dimensions H and L is simulated, the left side of the rectangle is clamped horizontally and the bottom side is clamped vertically.

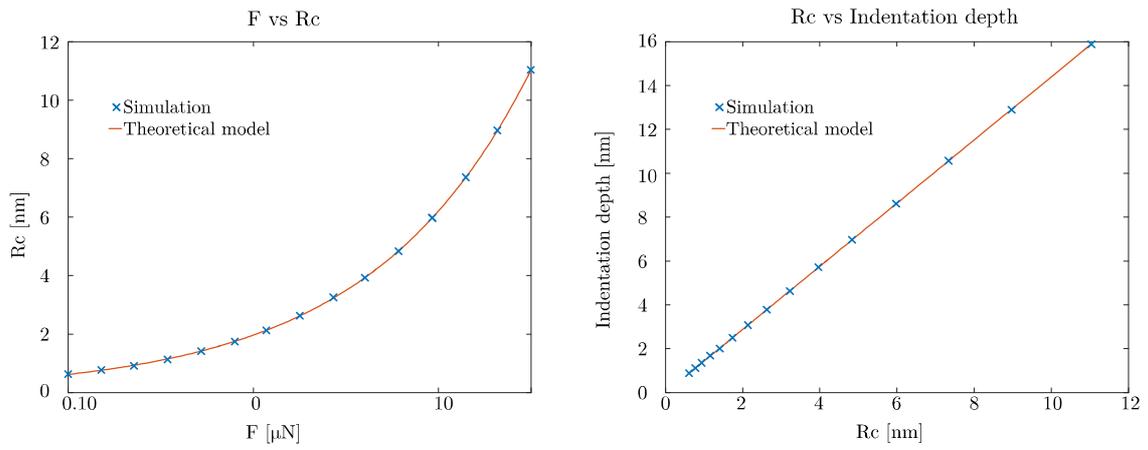

**Fig. S15. Validation of the contact model against analytical model for linear elasticity (vanishing $e, \mu, \kappa$ and $h$) (*31*).** Contact radius as a function of applied force (left) and indentation depth as a function of contact radius (right).

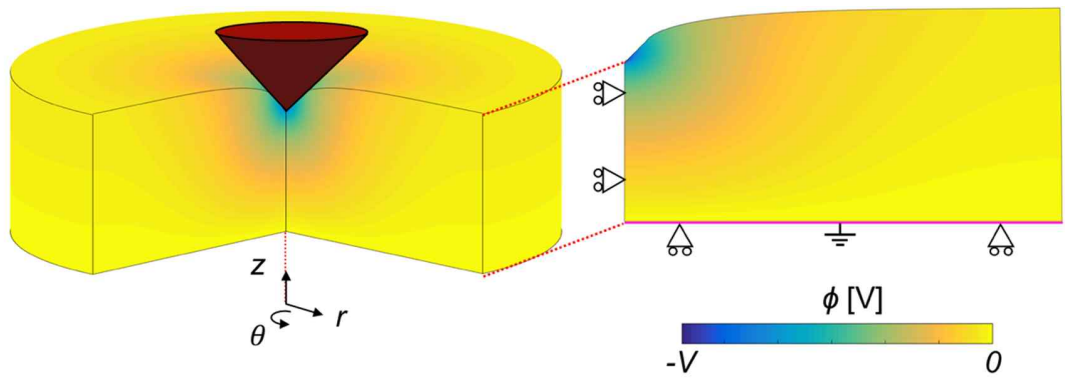

**Figure S16: Schematic of the axisymmetric three-dimensional model depicting the electric potential distribution upon indentation using a conical AFM probe.** In the two-dimensional axisymmetric model, the horizontal length is 40 nm and the vertical height is 20 nm.

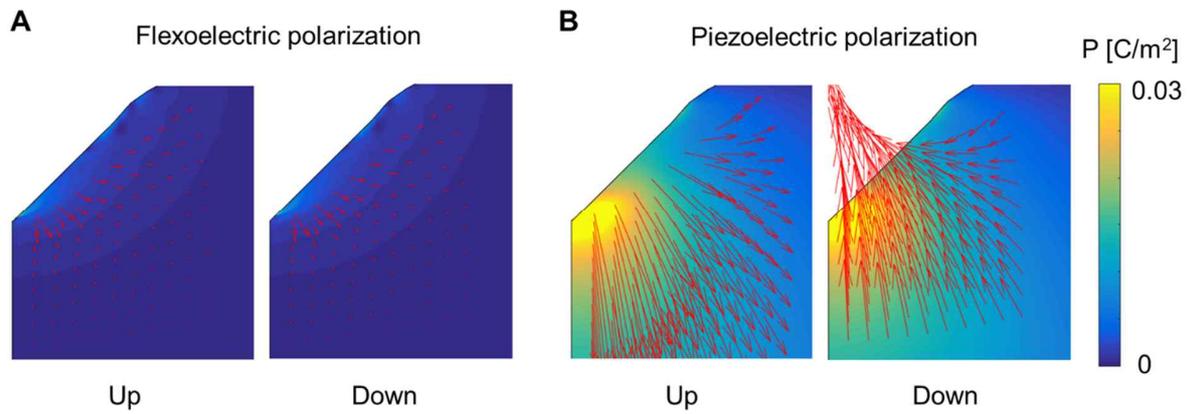

**Fig. S17**. **Polarization fields upon indentation induced by flexoelectricity and piezoelectricity for up and down ferroelectric domains, for $f = 10\ V$.** The polarization plotted here is relative to the remanent polarization.

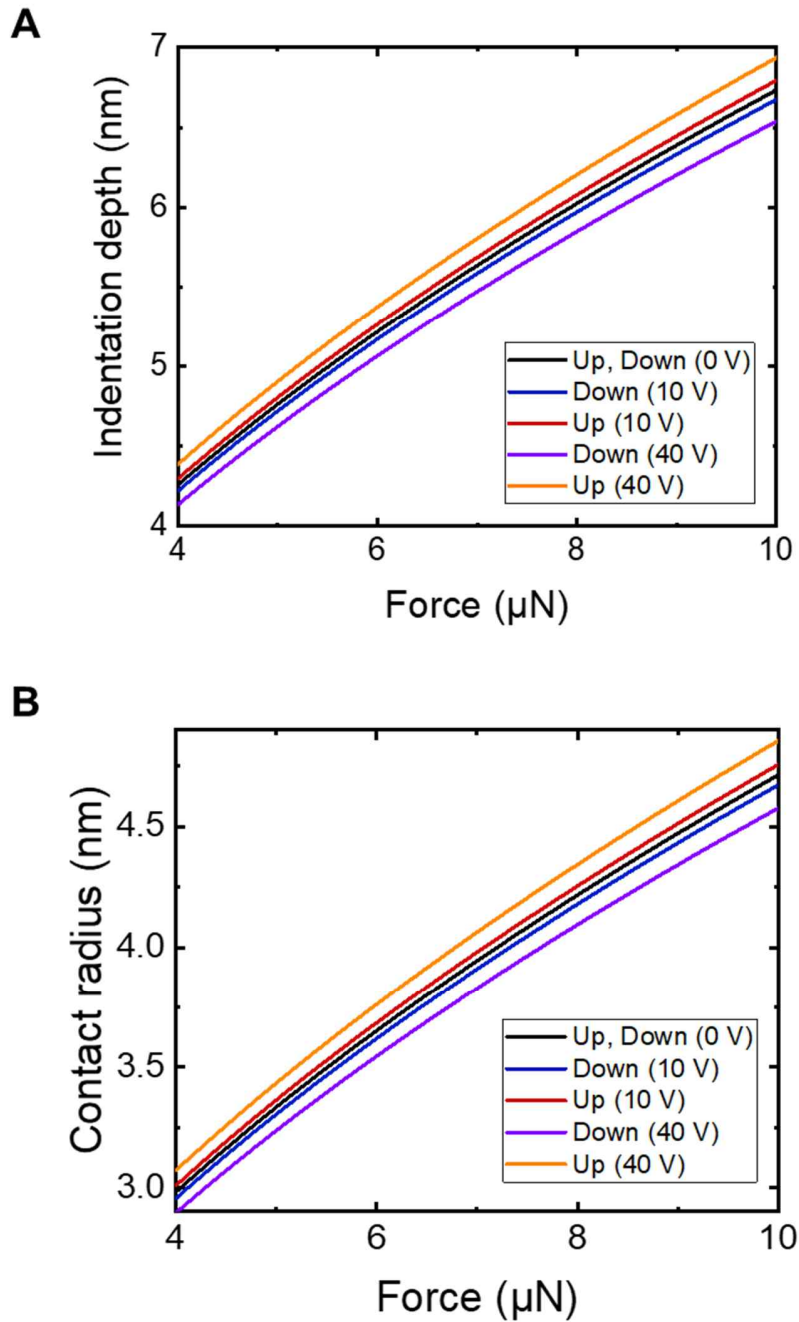

**Fig. S18. The effect of flexoelectricity on indentation depth and contact radius in up and down domains.** **(A)** Indentation depth and **(B)** contact radius as functions of applied force for up and down polarized domains for different values of the flexocoupling coefficient: $f = 0\,V$, $f = 10\,V$ and $f = 40\,V$. The observed asymmetry between up and down domains is stronger for larger $f$ and vanishes in the absence of flexoelectricity ($f = 0\,V$).

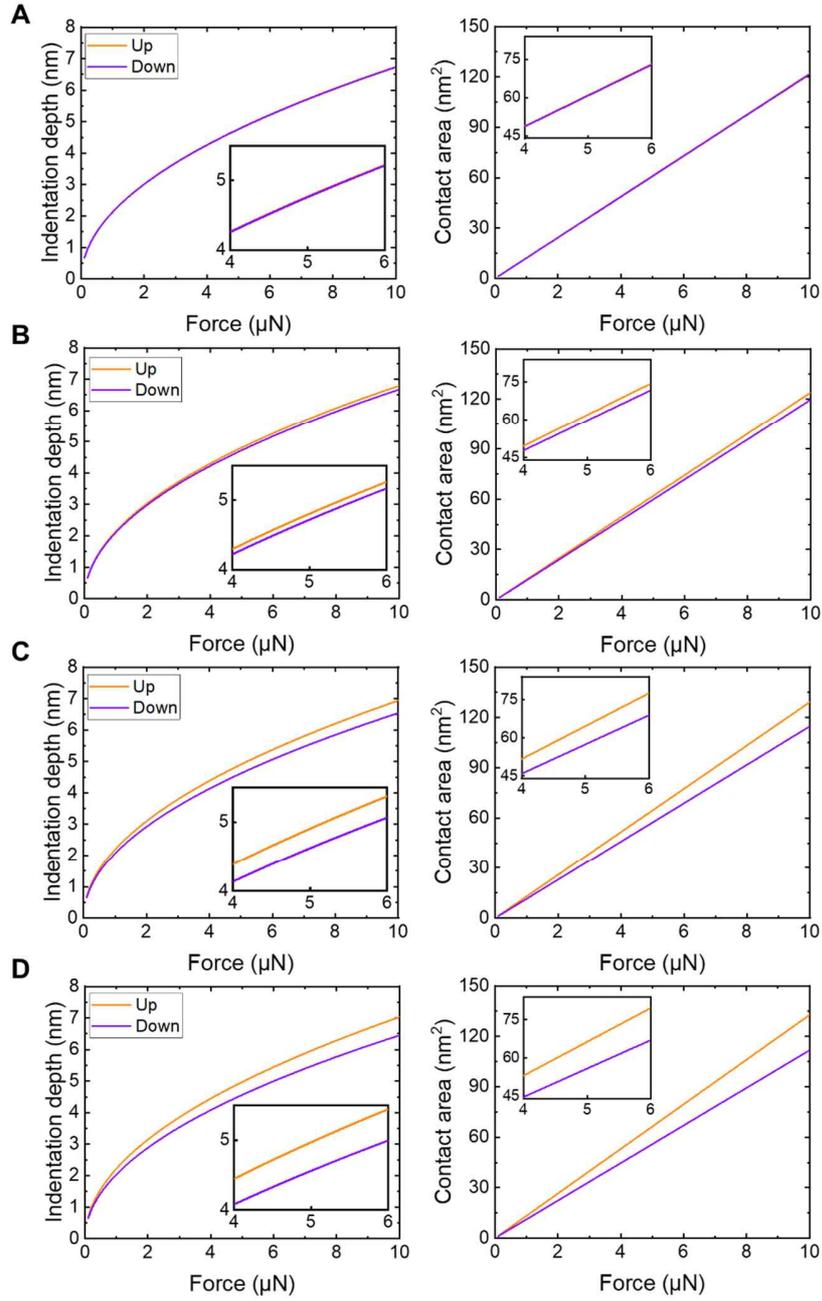

**Fig. S19. Asymmetric indentation depth and contact area of ferroelectric up (yellow) and down (purple) domains with different flexocoupling coefficients.** Indentation depth and contact area **(A)** $f = 1\,V$, **(B)** $f = 10\,V$, **(C)** $f = 40\,V$ and **(D)** $f = 54\,V$. Flexocoupling coefficients in Figs. S19C and S19D are based on experimental values in *(25)*.

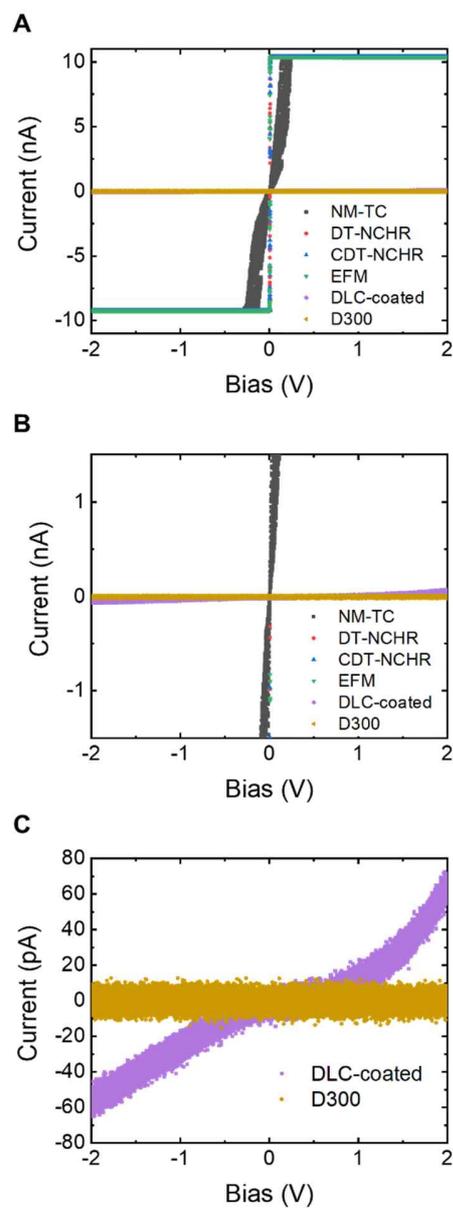

**Fig. S20**. *I-V* curve measurement on highly ordered pyrolytic graphite using six different probe types.

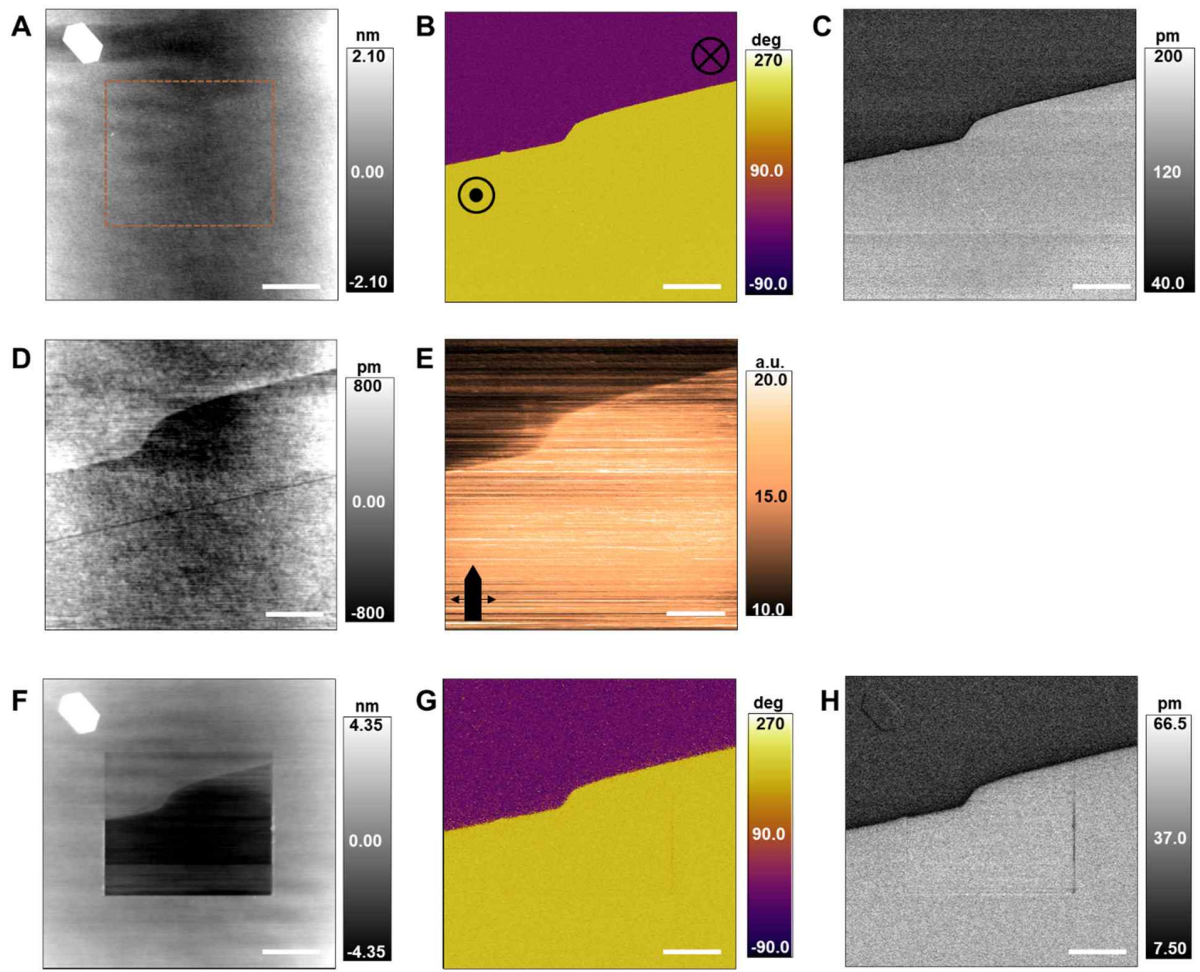

**Fig. S21. Asymmetric friction and wear of stoichiometric LiNbO₃ single crystal. (A)** Height, **(B)** PFM phase and **(C)** PFM amplitude of pristine surface of stoichiometric LiNbO₃. **(D)** Height, **(E)** friction during the milling scan. **(F)** Height, **(G)** PFM phase and **(H)** PFM amplitude after ten milling scans at a loading force of 5 μN and a scan rate of 1.95 Hz. All scans are measured at 20°C and 20% relative humidity.

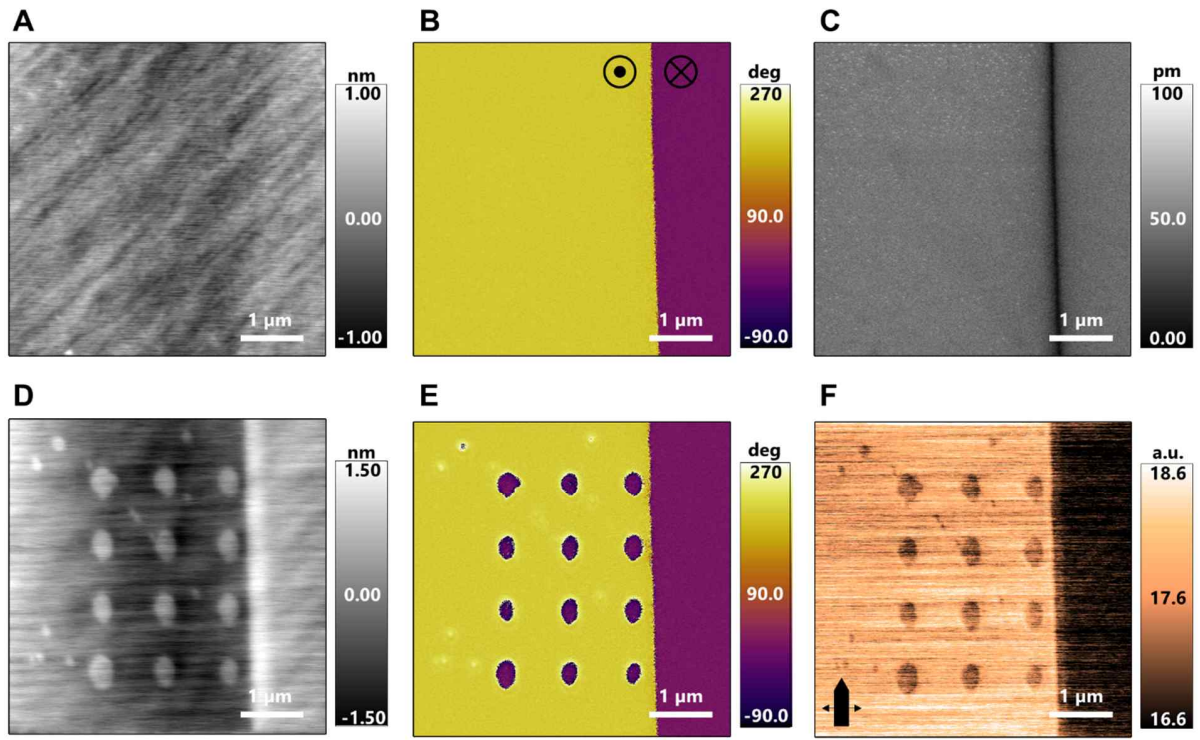

**Fig. S22. Switchable tribological asymmetry in PPLN. (A)** Height, **(B)** PFM phase and **(C)** PFM amplitude of pristine PPLN. **(D)** Height and **(E)** PFM phase after switching 3 × 4 arrays of down domains and ten milling scans at optimized condition. **(F)** Friction during the fifth milling scan. The friction of switched domains is lower than that up domains, and the resulting topography is also higher in the switched domains. All scans are measured at 20°C and 20% relative humidity.

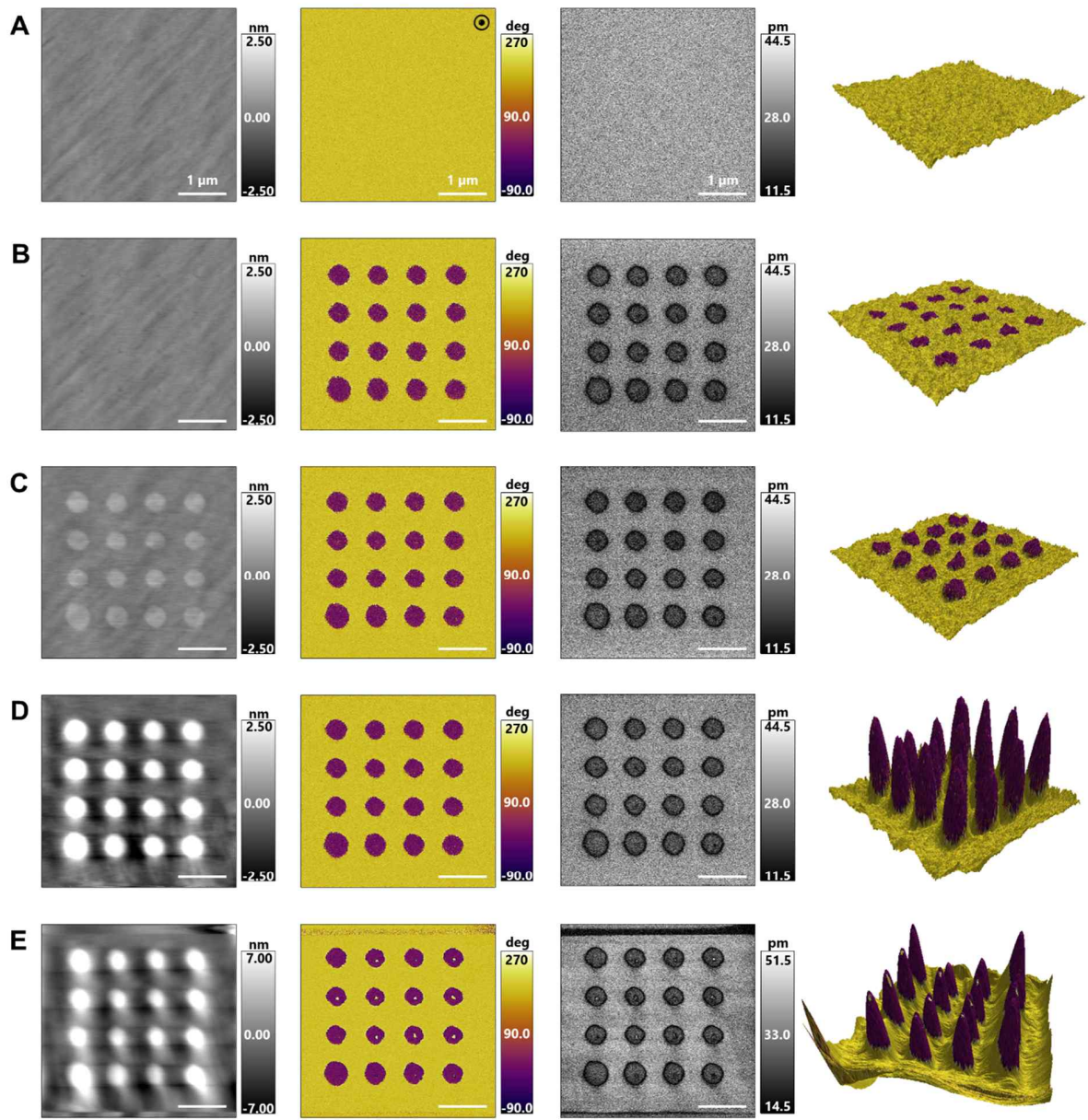

**Fig. S23**. **Fabrication of nanopillars in Fig. 3.** Height, PFM phase, PFM amplitude, and 3D image of height superimposed by PFM phase of **(A)** pristine up domain of PPLN, **(B)** after writing domains by electrical bias through the diamond probe, **(C)** after two millings, **(D)** after twenty millings, and **(E)** after sixty millings. Fig. S23E shows the existence of up domain at the core of the pillar, maybe because of the insufficient bias to switch the crystal, and it can be correlated with the saturation of the pillar growth. The z-scale of the 3D image is identical to the 2D height image. All scans are measured at 20°C and 20% relative humidity.

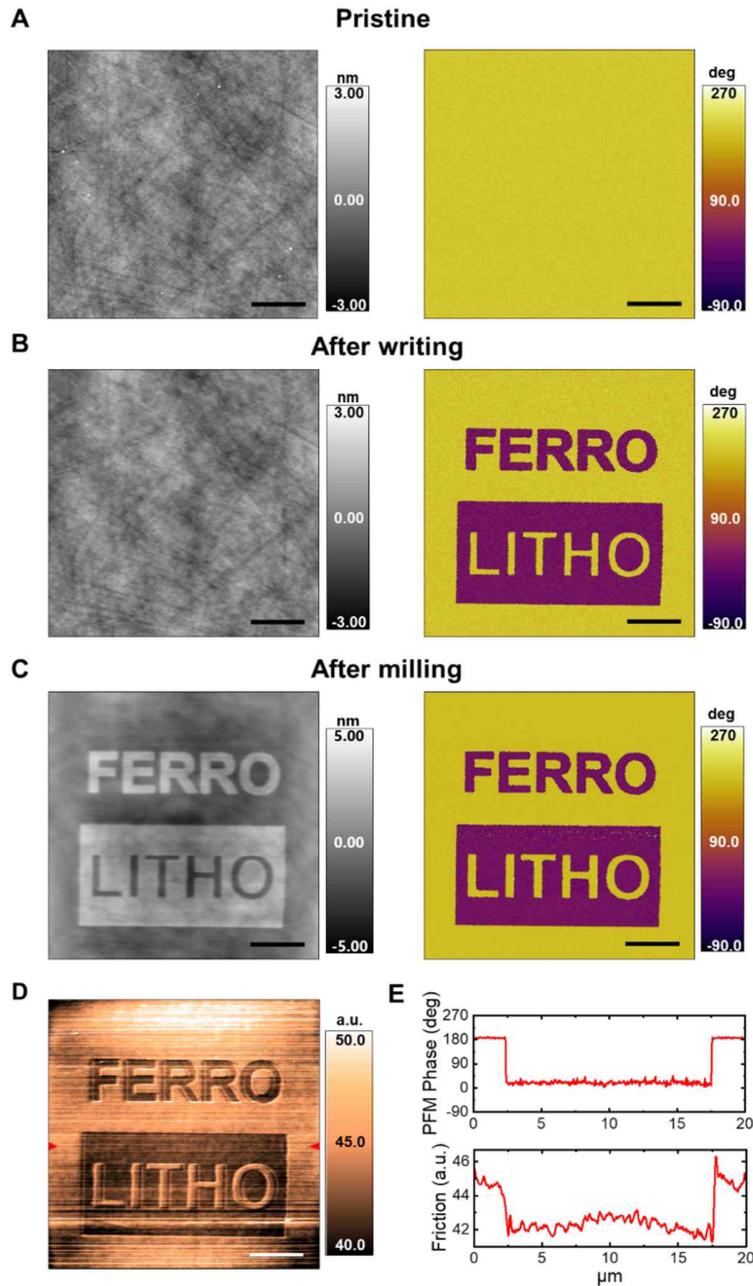

**Fig. S24: Nanostructuring using switchable wear asymmetry of ferroelectric LiNbO₃ thin film.** **(A)** Height and PFM phase of pristine LiNbO$_3$ thin film. **(B)** Height and PFM phase after the domain switching of "FERRO" and rectangular background of "LITHO" with down domains. **(C)** Height and PFM phase after multiple etching scans. **(D)** Friction during the milling scan showing higher friction in up domains and lower friction in down domains. **(E)** Line profiles of friction and PFM phase along the red markers in Fig. S24D. Line profile of PFM phase is obtained from Fig. S24C. All scans are measured at 20°C and 20% relative humidity. Scale bars are 4 μm.

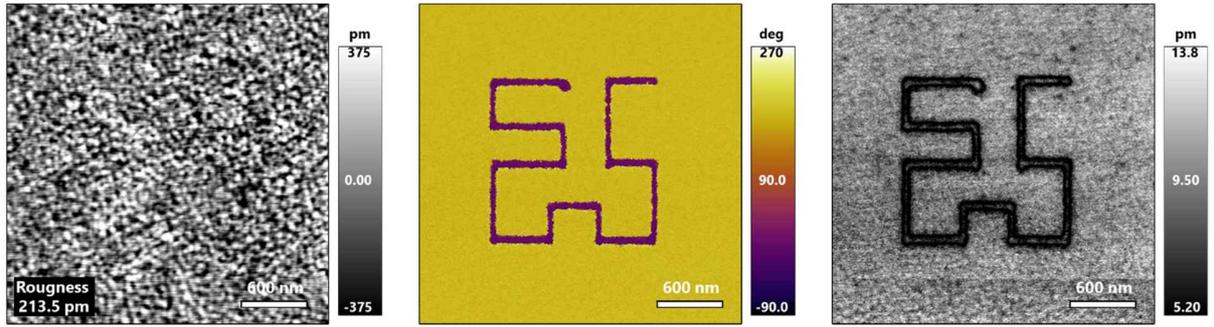

**Fig S25: Height, PFM phase and PFM amplitude images obtained after the electrical switching of up domains to down domains by applying 5 V to the AFM tip.**

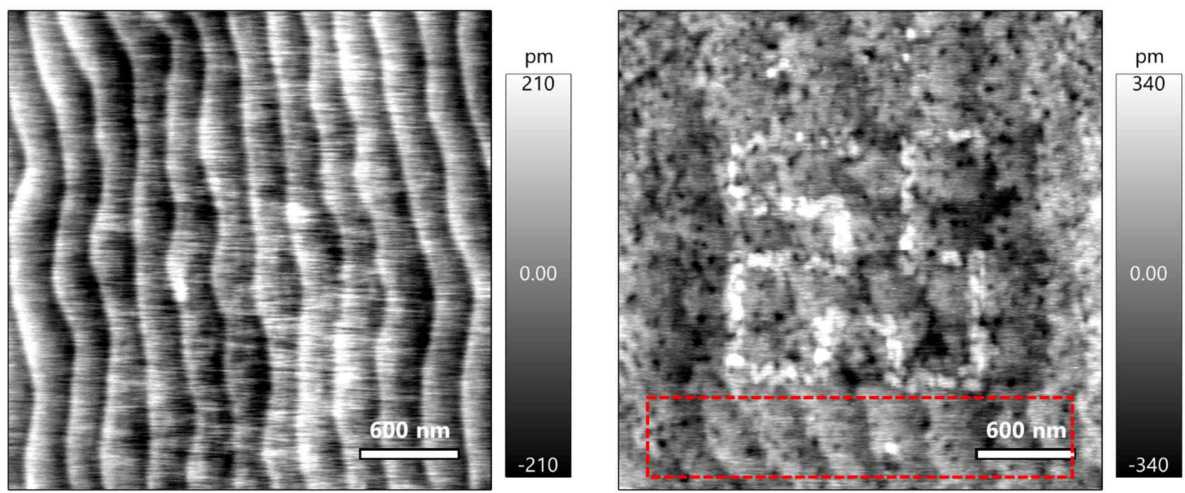

**Figure S26:** Surface topography of TiO$_2$ terminated (001)-oriented substrate SrTiO$_3$ (left) and etched PbTiO$_3$ thin film showing atomistic terrace edge features.

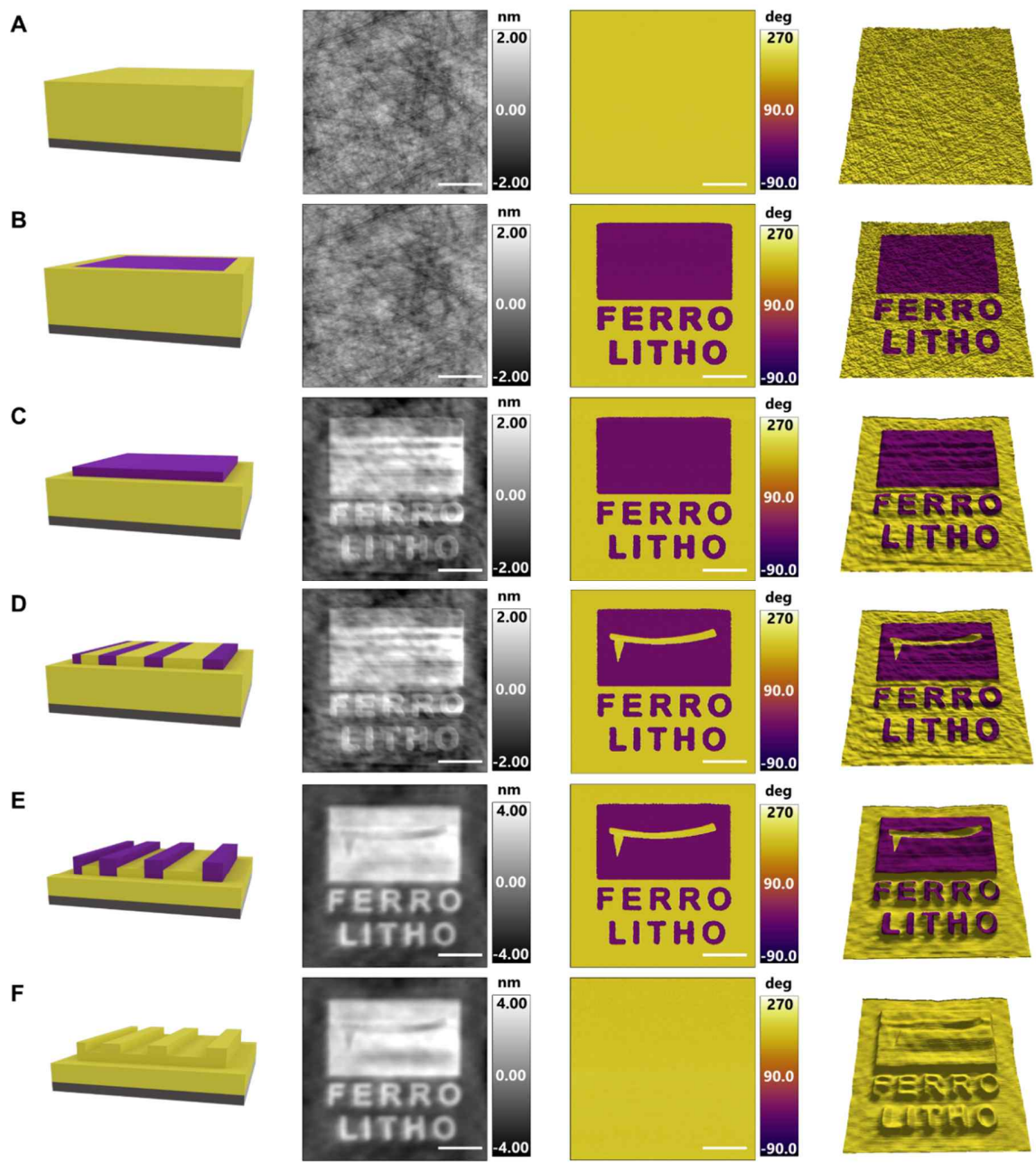

**Fig. S27. Demonstration of 3D nanostructure fabrication of ferroelectric LiNbO$_3$ thin film.** Schematic in each step, height, PFM phase and 3D surface superimposed with the PFM phase of **(A)** pristine, **(B)** after first polarization switching, **(C)** after first milling process, **(D)** after second polarization switching, **(E)** after second milling process and **(F)** after switching to uniform up polarization. Scale bars in height and PFM phase images are 3 μm. In 3D images, scan area is 12.5 μm × 12.5 μm and the z-scale is identical to 20 nm.

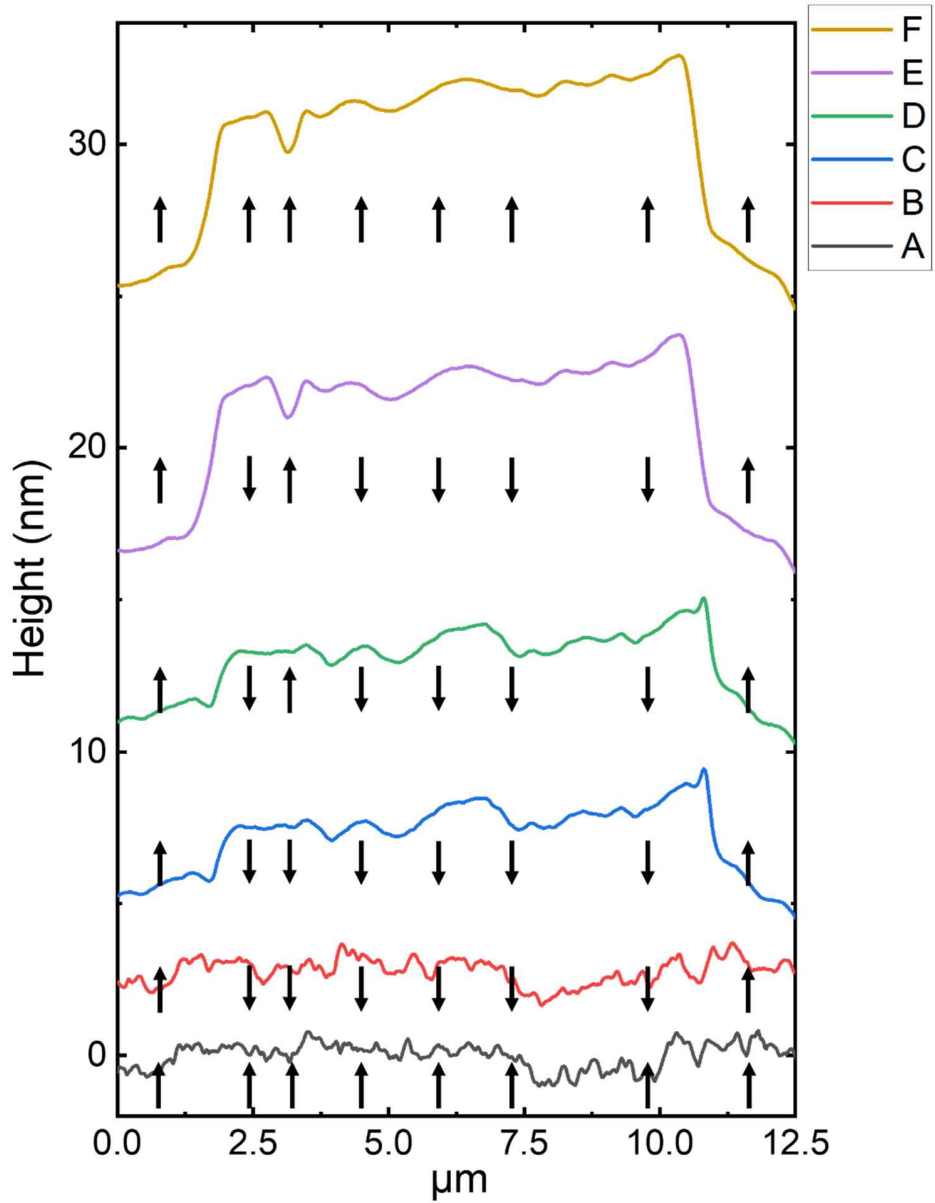

**Fig. S28. Structural evolution during the 3D nanostructuring in Fig S27**. The line profiles are obtained in the same position along with the AFM tip feature in Fig. 4B and arrows indicate the domain orientation. Line profiles of **(A)** pristine, **(B)** after 1st polarization switching, **(c)** after 1st milling process, **(d)** after 2nd polarization switching, **(e)** after 2nd milling process and **(f)** after switching to uniform up polarization.

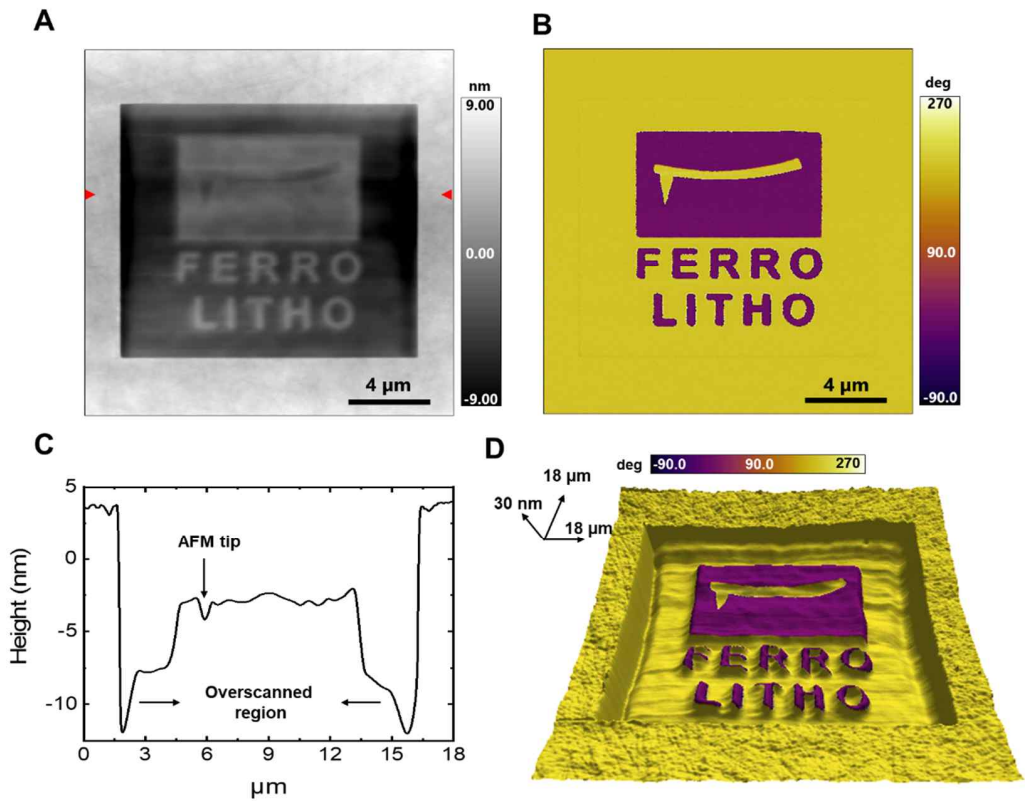

**Fig. S29.** **(A)** Height and **(B)** PFM phase with pristine background after 3D nanostructuring. **(C)** Line profile along with the AFM tip feature in Fig. S29A (red arrow). **(D)** 3D surface image with color overlapped with its PFM phase. The non-uniform height of the overscanned region originates from the distortion reduction mechanism during the contact imaging of AFM, also described in Fig. S3.

| Probe | Description | Spring constant (N/m, nominal) | Asymmetric etching of PPLN |
|---|---|---|---|
| NM-TC | Single crystal diamond (conductive) | 350 | O |
| DT-NCHR | Diamond-coated (conductive) | 80 | O |
| CDT-NCHR | Diamond-coated (conductive) | 80 | O |
| EFM | Pt/Ir-coated (conductive) | 2.8 | X |
| HQ:NSC16/HARD/ Al BS (DLC-coated) | Diamond-like-carbon-coated (weakly conductive) | 40 | O |
| D300 | Single crystal diamond (non-conductive) | 40 | O |

**Table S1. Probe selection for the asymmetric etching of PPLN.**

| $f$ [V] | $F_f^{up}/F_f^{down}$ |
|---|---|
| 1 | 1.0036 |
| 10 | 1.0361 |
| 40 | 1.1236 |
| 54 | 1.1881 |

**Table S2. Ratio of friction in up and down domains for different values of the flexocoupling coefficient.** Note that the flexoelectric tensors are isotropic (*63*). Isotropic tensors of 40 V and 54 V are based on experimental values in (*25*).